\documentclass[]{article}
\usepackage[T1]{fontenc}\usepackage[utf8]{inputenc}
\usepackage{amsmath,amsfonts,amsthm,amssymb,graphicx,longtable,color,subfigure,todonotes,afterpage,authblk,fullpage}
\usepackage{algorithm}\usepackage{algorithmicx,algpseudocode}
\title{Proliferation of Twinning in HCP Metals: \\ Application to Magnesium}
\author[1]{D. Sun \thanks{dingyi\_sun@brown.edu}}
\author[2]{M. Ponga \thanks{mponga@mech.ubc.ca}}
\author[3]{K. Bhattacharya\thanks{Corresponding Author: bhatta@caltech.edu}}
\author[3]{M. Ortiz \thanks{{ortiz@aero.caltech.edu}}}

\affil[1]{School of Engineering, Brown University, Providence, RI 02912, USA}
\affil[2]{Department of Mechanical Engineering, University of British Columbia, Vancouver, BC V6T 1Z4, Canada}
\affil[3]{Division of Engineering and Applied Sciences, California Institute of Technology, Pasadena, CA 91125, USA}

\begin{document}

\maketitle

\begin{abstract}
Plastic deformation of metallic alloys usually takes place through slip, but occasionally involves twinning.   In particular, twinning is important in hexagonal close packed materials where the easy slip systems are insufficient to accommodate arbitrary deformations.  While deformation by slip mechanisms is reasonably well understood, less remains known about deformation by twinning.   Indeed, the identification of relevant twinning modes remains an art.  In this paper, we develop an universal framework combining fundamental kinematic definition of twins with large scale atomistic calculations to predict twinning modes of crystalline materials.  We apply this framework to magnesium where there are two accepted twin modes -- tension and compression, but a number of anomalous observations.  Surprisingly, our framework shows that there are a very large number of twinning modes that are important in the deformation process of magnesium consistent with the anomalous observations.  Thus, in contrast to the traditional view where plastic deformation is kinematically partitioned between a few modes, our result argues that the physics of deformation in HCP materials is governed by an energetic and kinetic competition between numerous possibilities. Consequently, our findings suggest that the commonly used models of deformation physics need to be revisited in order to take into account a broader and richer variety of twin modes, and potentially points to new avenues of improving the mechanical properties.
\end{abstract}

\section{Introduction}\label{sec:Intro}

Magnesium alloys have amongst the highest strength to weight ratio (with a density of 1.8 gm/cm$^3$ and yield strength exceeding 100MPa) of known metals, and have been explored for automotive, biomedical and other engineering applications.  However, these alloys often have limited ductility and suffer sudden, almost brittle, failure.  We refer the reader to recent reviews \cite{JoostKrajewski2017,XianhuaYuxiaoFusheng2016,KusnierczykBasista2017,Kulekci2008} and to \cite{ChristianMahajan1995} for additional background on deformation twinning. 

\begin{figure}
\centering
\includegraphics[width=0.8\linewidth]{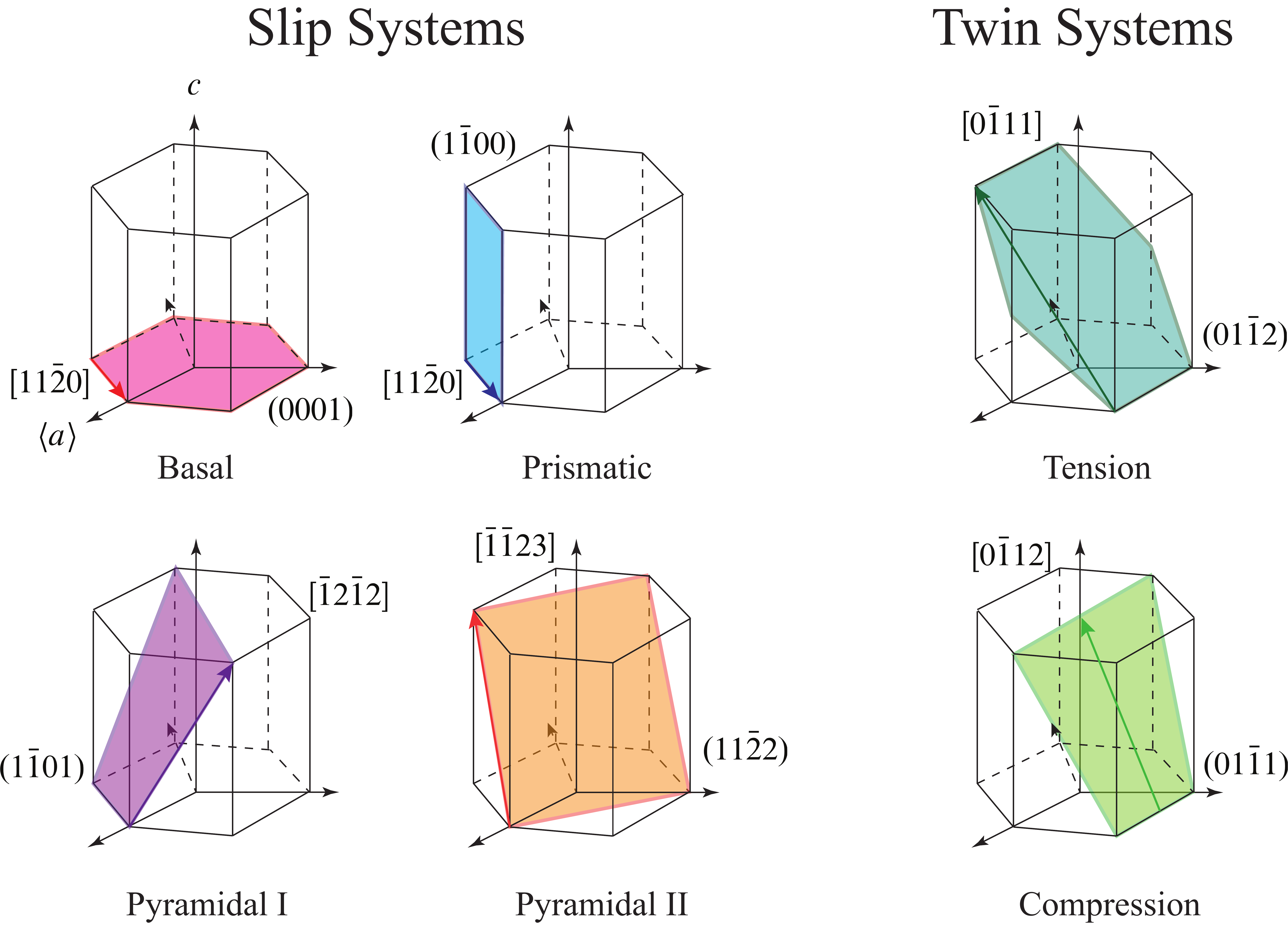}
\caption{Mechanisms of deformation in HCP materials}
\label{fig:HCPMechanisms}
\end{figure}

The high strength, as well as the limited ductility, has its origins in the hexagonal closed packed crystal structure. It is well known that magnesium like most HCP materials has an easy slip along the so-called basal planes (see Figure \ref{fig:HCPMechanisms}). However, slip is significantly more difficult along any plane that is not the basal plane, including the so-called prismatic and pyramidal planes, where the stress required to slip the material is approximately one and two orders of magnitude larger than the basal plane, for the prismatic and pyramidal slip systems. Thus, deformation of these materials also involves twinning where a portion of the crystal undergoes a shear that restores the lattice.  The twinned and untwinned regions are separated by a twin plane and the two sides possess either an identical crystallographic plane or an identical crystallographic direction.   We refer the reader to Christian and Mahajan\cite{ChristianMahajan1995} for a comprehensive introduction to deformation twinning.

It has long been known \cite{ChristianMahajan1995,Barnett2007a} that the $\{10\bar{1}2\}\langle10\bar{1}1\rangle$ tension twin plays an important role in the deformation of magnesium and other hexagonal close packed materials.  The twin plane as well as the twinning direction of this twinning mode are shown in Figure \ref{fig:HCPMechanisms}.  It is now also accepted that the $\{10\bar{1}1\}\langle10\bar{1}2\rangle$ compression twin is also an important deformation mechanism \cite{Barnett2007b}.  These two twinning modes as well as the four slip modes shown in Figure 1 are the basis of a number of models of deformation in magnesium alloys \cite{StaroselskyAnand2003,BrownEtAl2005,ChangKochmann2015,ChangEtAl2017}.

An understanding of the deformation modes are critical for the improvement for the mechanical properties of magnesium alloys for many reasons.  The key to improving the properties is to reduce the high anisotropy created by the contrast in strength between the basal mode and the other modes.  In other words, one needs not only to strengthen the basal mode through precipitate and solute hardening, but also soften the other modes by chemical means.  Further, texture is also critical for improved ductility in commercial materials, and this is obtained by deformation processes \cite{AgnewEtAl2004,BrownEtAl2005,XiaEtAl2005,YamashitaEtAl2001}.  

For this reason, twinning in magnesium alloys has been the focus of much recent activity.  Much of this work has focussed on the tension twins.   El Kadiri \emph{et al.} \cite{TschoppEtAl2013} as well as Li and Ma \cite{LiMa2009}) have studied the detailed atomistic shuffle mechanism related to these twins.  Beyerlein and coworkers \cite{BeyerleinMcCabeTome2011a,BeyerleinMcCabeTome2011b,WangBeyerleinHirth2012,WangEtAl2011} have studied the nucleation and growth process including the role of particular twinning dislocations and the role of grain boundaries.  Others  \cite{MorrisYeYoo2005,WangEtAl2010} have used first principles to study of twin boundary and stacking fault energies, and how they are affected by the addition of rare earth elements.

Still, all this work in the literature is premised on the fact that twinning is limited to the tension and compression systems shown in Figure \ref{fig:HCPMechanisms}.  However, there has long been reports of observations of other twinning modes in magnesium and its alloys.  Various authors have reported a $\{10\bar{1}3 \}\langle30\bar{3}2 \rangle$ system \cite{BrownEtAl2005,JiangEtAl2006,KitaharaEtAl2007,ReedHill1960,WangEtAl2011,WangBeyerleinHirth2012}.  Reed-Hill \cite{ReedHill1960} observed twinning on an irrational twin plane.  Brown {\it et al.} \cite{BrownEtAl2005} observed anomalous peaks in their study of twinning using neutron diffraction peaks.  Molodov \emph{et al.} \cite{MolodovEtAl2016} also discuss the discovery of some twin modes whose orientations do not match the classical modes. Jiang {\it et al.} \cite{JiangEtAl2006} discuss double twinning modes.  Liu {\it et al.} \cite{LiuEtAl2014} report a "twinning like lattice reorientation"  across a faceted plane.  Indeed, Christian and Mahajan \cite{ChristianMahajan1995} have an extensive discussion of "anomalous" twinning modes in magnesium and other HCP metals, some of which may be observed additionally in \cite{YoshinagaObaraMorozumi1973,KitaharaEtAl2007,MolodovEtAl2016,YoshinagaHoriuchi1963}.

In this work, we exploit the availability of computational power and atomistic methods to conduct a systematic and extensive search of all possible twinning modes.  We start from the fundamental kinematic definition of a twin following Cahn, Bilby and Crocker and others \cite{Cahn1954,BilbyCrocker1965} and the mathematical formulation of Ericksen, Pitteri, Zanzotto and others \cite{PitteriZanzotto2003} in Section \ref{sec:kinematics}.   Unlike the traditional approach of using this definition to look for particular modes guided by experimental observation, we use the formulation to generate an extensive list of kinematically admissible twin modes.  

We examine the energy landscape of the kinematically admissible twin modes using atomistic methods in Section \ref{sec:ener}.  We use the  modified MEAM potential of Wu \emph{et al.} \cite{WuFrancisCurtin2015} in this work, to compute the twin boundary energy as well as the energy barrier for twinning.  We compute the kinetic rates for the various twinning modes following the ideas of Weiner \cite{Weiner2002} in Section \ref{sec:kinetic}.  Finally, we compute the yield surface in Section \ref{sec:yield}.

Surprisingly, we find that there are a very large number of twinning modes that are important in the deformation process of magnesium.  Indeed, there are a very large number of modes with kinematic, energetic and rate attributes comparable to those of the accepted tension and compression twin modes.  Indeed, as many as twelve twin modes are relevant to the yield behavior, and there are still other modes which are very close.  Some of our modes are consistent with the anomalous observations.

%%%%%%%%%%%%%%%%%%%%%%%%%%%%%%%%
%%%%%%%%%%%%%%%%%%%%%%%%%%%%%%%%
\section{Kinematics of twins} 
\label{sec:kinematics}

%%%%%%%%%%%%%%%%%%%%%%%%%%%%%%%%
\subsection{Crystal}

A crystal is a periodic arrangement of atoms or discrete points.  The simplest crystal is a Bravais lattice where we have only one atom per unit cell:
\begin{equation}\label{eq:Bravais}
\mathcal{L}_b=\left\{\mathbf{x}:\mathbf{x}=\sum_{i=1}^3 \nu^i\mathbf{e}_i, \nu^i \text{ integers} \right\}
\end{equation}
where the lattice vectors $\{\mathbf{e}_i\}_{i = 1}^3$ are linearly independent and describe the periodicity or the unit cell of the lattice.  Note that the choice of lattice vectors is not unique.  Indeed, two sets of linearly independent vectors $\{\mathbf{e}_i\}_{i = 1}^3$ and $\{\mathbf{f}_i\}_{i = 1}^3$ generate the same lattice if and only if 
\begin{equation} \label{eq:muij}
\mathbf{f}_i = \mu_i^{\ j} \mathbf{e}_j
\end{equation}
for some $\{\mu_i^{\ j}\}_{i,j = 1}^3$ a $3x3$ matrix of integers with determinant $\pm 1$.  We denote $\mathcal{M}$ to be the set of all $3x3$ matrix of integers with unit determinant.

Not all crystals are Bravais lattices.  Indeed, a hexagonal closed packed crystal is not a Bravais lattice.  However,  any crystal can be expressed as a multi-lattice or a finite collection of identical Bravais lattices which are translated related to each other:
\begin{equation}\label{eq:NonBravais}
\mathcal{L}_{nb}=\left\{\mathbf{x}:\mathbf{x}=\sum_{i=1}^3 \nu^i\mathbf{e}_i+\sum_{k=1}^{K-1}\eta_k\mathbf{p}_k, \nu^i \text{ and }\eta_k\text{ integers} \right\}
\end{equation}
where the lattice vectors $\{\mathbf{e}_i\}_{i = 1}^3$ are linearly independent and describe the periodicity of the lattice as before, and the shifts $\{\mathbf{p}_k\}_{k=1}^K$ are vectors that describe the translation of the constituent Bravais lattices relative to each other.  The choice of lattice vectors and shifts is not unique for a given 
crystal.  The exact necessary and sufficient conditions are somewhat involved; however, the lattice vectors still satisfy (\ref{eq:muij}) \cite{PitteriZanzotto2003}.

We note a further degeneracy in our description of crystals.  For any crystal there is a minimal unit cell involving the smallest number of atoms required to describe the periodicity.  However, we can take a supercell consisting of as many multiples of the unit cell as we choose, and use this as the unit cell.  In other words we can take $K$ to be as large as possible. By this act, we allow for shuffling to take effect and act on the motion of the atoms in the configuration. Consequently, we are adding twin modes with shuffles. We refer the reader to \cite{PitteriZanzotto2003} for additional details and discussion. 

%%%%%%%%%%%%%%%%%%%%%%%%%%%%%%%%
\subsection{Twin}
\begin{figure}
\centering
\includegraphics[width=0.5\linewidth]{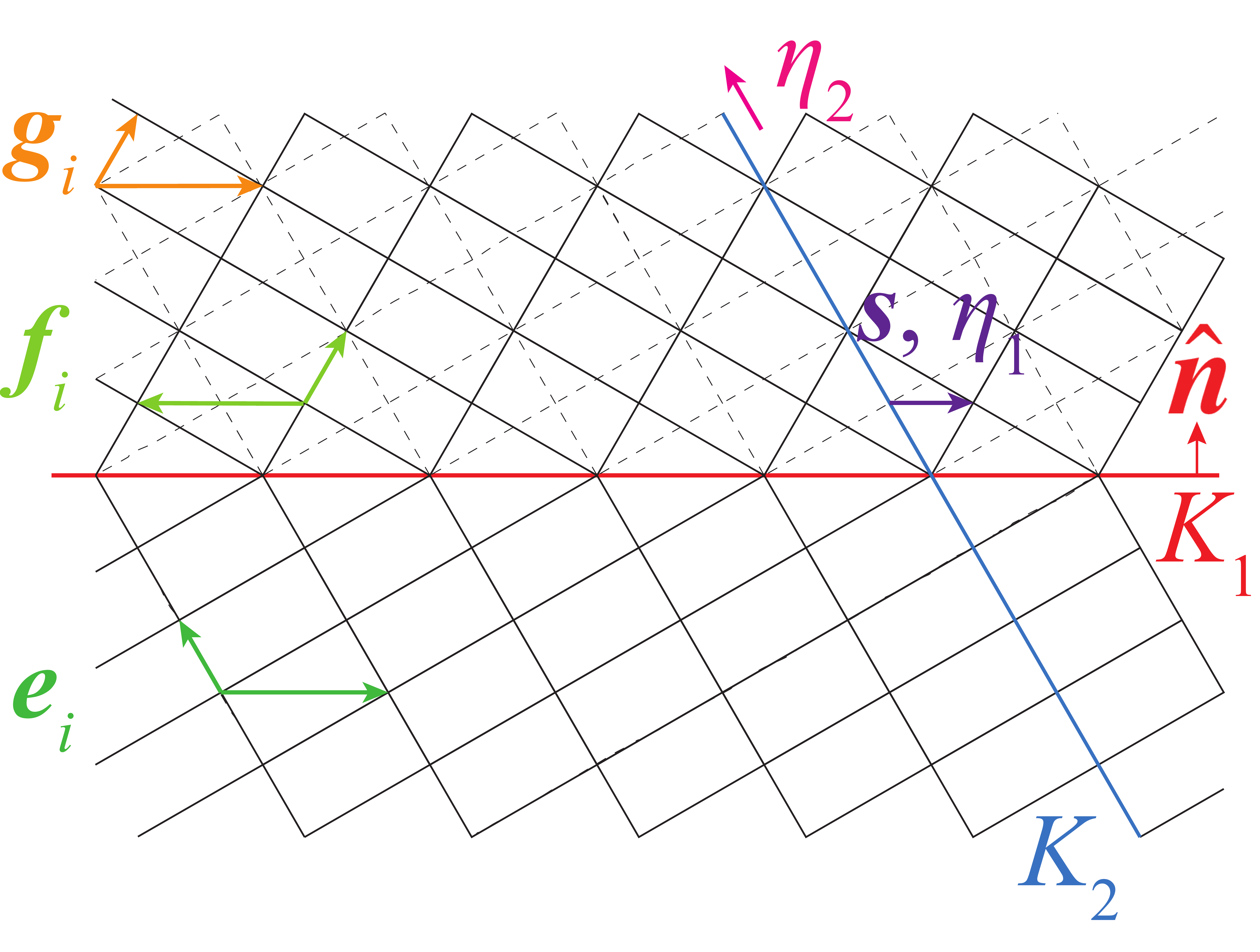}
\caption{A schematic example of a twin.}
\label{fig:twin}
\end{figure}

A twin is a planar discontinuity in a crystal where the lattice on one side may be described both
\begin{itemize}
\item as a rotation and
\item as a simple shear
\end{itemize}
of the lattice on the other \cite{Cahn1954,BilbyCrocker1965,ChristianMahajan1995,PitteriZanzotto2003}\footnote{The traditional definition requires that the two sides either have a common crystallographic plane or crystallographic direction, but can be shown to be equivalent to the second requirement.}.  This is shown schematically in Figure \ref{fig:twin}.  
The twin plane is typically denoted by $K_1$, the direction of shear as $\eta_1$ and the magnitude of shear as $s$.  One can show that there is another undistorted plane denoted as $K_2$ and an undistorted direction $\eta_2$.

We now seek to study the implications of this definition.  Let $\{\mathbf{e}_i\}$, denote the lattice vector on one side of the material.  Since the lattice on the other side can be obtained by a rotation, we must have a set of lattice vectors $\{\mathbf{f}_i\}$ that describe the lattice on the other side such that
\begin{equation}\label{eq:fi}
\mathbf{f}_i=\mathbf{Qe}_i,
\end{equation}
for a rotation $\mathbf{Q}$, i.e. $\mathbf{Q^\mathrm{T}Q=QQ^\mathrm{T}=I}$ (with $\mathbf{I}$ denoting the identity ) and $|\det\mathbf{Q}|=1$.   Further, since the lattice on the other side can be obtained by a simple shear, we must have a set of lattice vectors $\{\mathbf{g}_i\}$ that describe the lattice on the other side such that
\begin{equation}\label{eq:gi}
\mathbf{g}_i=(\mathbf{I+s\otimes \hat{n}})\mathbf{e}_i.
\end{equation}
for some non-zero vector $\mathbf{s}$ where $\hat{\mathbf{n}}$ denotes the normal to the twin plane.  Since 
$\{\mathbf{f}_i\}$ and $\{\mathbf{g}_i\}$ generate the same lattice, they must be related by 
\begin{equation}\label{eq:gf}
\mathbf{g}_i=\mu_i^{\ j} \mathbf{f}_j,
\end{equation}
for some $\{\mu_i^{\ j}\} \in \mathcal{M}$. Equivalently, 
\begin{equation}\label{eq:gfExpanded}
(\mathbf{I+s\otimes\hat{n}})\mathbf{e}_i=\mu_i^{\ j}\mathbf{Qe}_j.
\end{equation}
Note that $K_1$ is the plane with normal $\hat{\mathbf{n}}$, $\eta_1$ is the direction of  vector $\mathbf{s}$ and the magnitude of shear $s$ is the magnitude of the vector $\mathbf{s}$.  A solution with $K_1$ rational (i.e, the vector $\hat{\mathbf{n}}$ has components whose ratios are rational numbers in some reciprocal lattice basis $\{\mathbf{e}^i\}$) is known as a type I twin while the solution with $\eta_1$ rational (i.e, the vector $\mathbf{s}$ has components whose ratios are rational numbers in some  lattice basis $\{\mathbf{e}_i\}$)  is known as a type II twin.  A solution where both $K_1$ and $\eta_1$ are rational is known as a compound twin.

Note that in our description above, we only require the unit vectors to satisfy the condition (\ref{eq:gfExpanded}).  Thus the shift vectors on one side may not related to a shift vectors on the other side by a simple shear.  This is often described as a shuffle, i.e., the atoms with the unit cell are not convected to a rotation-related position by the simple shear, but then shuffle to the rotation-related position.  Such shuffles only displacement within the unit cell and do not cause any macroscopic change to the lattice.

This equation has been widely used to describe various twinning modes by showing that there are rotations and matrices $\mu_i^{\ j}$ consistent with this equation.

%%%%%%%%%%%%%%%%%%%%%%%%%%%%%%%%
\subsection{Kinematically allowed twinning modes}

In this work, we turn the practice around and seek to use the twinning equation (\ref{eq:gfExpanded}) to generate an extensive list of kinematically allowable twinning modes.  In other words, we scan over descriptions of the lattice, i.e., lattice vectors $\{\mathbf{e}_i\}$ and matrices  $\mu_i^{\ j} \in \mathcal{M}$ and examine whether the twinning equation (\ref{eq:gfExpanded}) has a solution $\mathbf{Q}, \mathbf{s}, \hat{\mathbf{n}}$.  Each such solution describes a possible twinning mode. 

It is convenient to rewrite (\ref{eq:gfExpanded}).  For any set of linearly independent vectors $\{\mathbf{e}_i\}$, we introduce the reciprocal vectors $\{\mathbf{e}^i\}$ such that $\mathbf{e}_i \cdot \mathbf{e}^j = \delta_i^j$.  For any $\mu_i^{\ j} \in \mathcal{M}$, we introduce the tensor
\begin{equation}\label{eq:H}
\mathbf{H}=\mu_i^{\ j} \mathbf{e}_j\otimes \mathbf{e}^i.
\end{equation}
The twinning equation (\ref{eq:gfExpanded}) can now be rewritten as 
\begin{equation}\label{eq:DeformationGradient}
\mathbf{QH}=\mathbf{I+s\otimes \hat{n}}.
\end{equation}

This gives rise to the following problem: given $\mathbf{H}$, can we  $\mathbf{Q}, \mathbf{s}, \hat{\mathbf{n}}$ to satisfy (\ref{eq:DeformationGradient}).   Ball and James \cite{BallJames1987} provides an answer to this question.  The results states that (\ref{eq:DeformationGradient}) has a solution if and only if the tensor 
$\mathbf{C} = \mathbf{H}^T\mathbf{H} \ne {\mathbf I}$ has eigenvalues $\{\lambda_i\}_{i=1}^3$ that satisfy
\begin{equation}\label{eq:EigenvaluesC}
0< \lambda_1\le \lambda_2=1\le \lambda_3.
\end{equation}
Further, if this condition is satisfied, there are exactly two solutions and they are given by 
\begin{subequations}
\begin{gather}
\mathbf{s}=\rho \left(\sqrt{\dfrac{\lambda_3(1-\lambda_1)}{\lambda_3-\lambda_1}}\hat{\mathbf{v}}_1+\kappa\sqrt{\dfrac{\lambda_1(\lambda_3-1)}{\lambda_3-\lambda_1}}\hat{\mathbf{v}}_3 \right), \label{eq:TwinningShear}\\
\hat{\mathbf{n}}=\dfrac{1}{\rho}\left(\dfrac{\sqrt{\lambda_3}-\sqrt{\lambda_1}}{\sqrt{\lambda_3-\lambda_1}}(-\sqrt{1-\lambda_3}\hat{\mathbf{v}}_1+\kappa\sqrt{\lambda_3-1}\hat{\mathbf{v}}_3) \right), \label{eq:TwinningNormal} \\
\mathbf{Q}=(\mathbf{I+s\otimes\hat{n}})\mathbf{H}^{-1} \label{eq:Q}
\end{gather}
\end{subequations}
where $\kappa = \pm 1$, $\rho\ne 0$ is a normalization constant to ensure that the twin normal $\hat{\mathbf{n}}$ maintains a unit magnitude, $\{\hat{\mathbf{v}}_i\}_{i=1}^3$ are the (orthonormal) eigenvectors of $\mathbf{C}$ corresponding to the $\{\lambda_i\}$.

We can rule out the case $\mathbf{H}^T\mathbf{H} = {\mathbf I}$.  This condition means that ${\mathbf H}$ is a rotation and thus we have no discontinuity.  Thus, given $\mathbf{H}$, we either have no twinning mode, or exactly two conjugate twinning modes that share the same magnitude of shear but interchange $(K_1,\eta_1)$ and $(K_2,\eta_2)$.  

It is customary in the study of deformation twinning to confine $\mathbf{Q}$ to two-fold rotations, or rotations through $180^\circ$.  We do not impose any such restriction {\it a priori}.  We define the angle of rotation to be $\theta$ and can easily compute it as
\begin{equation}
\mbox{tr } \mathbf{Q} = 1 + 2 \cos \theta.
\end{equation}

Finally, note that there are a countably infinite choices of the matrix $\mu_i^{\ j} \in {\mathcal M}$ and a countably infinite choices of super cells for any given crystal.   So, it is not feasible to scan over all possible matrices and super cells.  However, we can use (\ref{eq:DeformationGradient}) and (\ref{eq:H}) to show that
\begin{equation}
\mu_i^{\ j} \mu_i^{\ j} = \mbox{tr } {\mathbf H}^T{\mathbf H} = 3 + |\mathbf{s}|^2.
\end{equation}
Thus, as the elements of $\mu_i^{\ j}$ become larger, the shear magnitude becomes larger.  As we shall see later, the amount of shear is generally an estimate of the energy barrier, and thus one only sees twinning modes with the smallest shear.  Therefore, it suffices to confine search scan to reasonable values of $\mu_i^{\ j}$.  Similarly, the number and magnitude of shuffles increases as one increases the size of the super cell.  This again increases the energetic barrier required for twinning.  Therefore, it is sufficient to confine the search to a reasonable number of supercells.

We close with a final comment regarding the twinning shear.  It is a common misconception in the literature that the atomic positions completely determine all the twinning elements, though experts have pointed out that this is not the case as the magnitude of shear is not determined (see \cite{PitteriZanzotto2003}).  In other words, one can have two twinning modes which share the same twin plane and shear direction, but different amounts of shear.    This can be intuitively clear by examination of Figure \ref{fig:twin}: notice that it is possible to restore all the atomic positions by shearing the top half of the crystal in the direction $\eta_1$ by a specific amount.   Mathematically, consider a type I twin where $\mathbf{Q}$ is a two fold rotation about a rational twin plane $\hat{\mathbf n}$ (i.e., $\mathbf{Q} = - \mathbf{I} + 2 \hat{\mathbf n} \otimes \hat{\mathbf n}$).  Because the twin plane is rational, we can choose our lattice vectors such that $\mathbf{e}_1, \mathbf{e}_2$ lie on the twin plane (i.e., $\hat{\mathbf n} \cdot \mathbf{e}_1 = \hat{\mathbf n} \cdot \mathbf{e}_2 = 0$).  It is easy to verify that the following solve the twinning equation (\ref{eq:gfExpanded}):
\begin{equation}
\mu_i^{\ j}=\left(\begin{matrix}
-1 & 0 & 0\\
0 & -1 & 0\\
\alpha & \beta & 0
\end{matrix} \right), \quad
\mathbf{s} = \frac{1}{\hat{\mathbf n} \cdot \mathbf{e}_3} (- \alpha \mathbf{e}_1 - \beta \mathbf{e}_2 )
\end{equation}
for any integers $\alpha, \beta$.  Consider the family of twinning modes with $\beta = 0$.  Note that they all share the same twin plane $\hat{\mathbf n}$ and twinning direction ${\mathbf e}_1/|{\mathbf e}_1|$, but differ by the magnitude of shear $\alpha |{\mathbf e}_1|/ (\hat{\mathbf n} \cdot \mathbf{e}_3)$.

This observation has the important consequence that diffraction techniques (x-ray diffraction or electron back-scatter diffraction) alone can {\it not} determine the twinning shear and one needs to measure the twinning shear by measuring macroscopic strain.

%%%%%%%%%%%%%%%%%%%%%%%%%%%%%%%%
\subsection{Application to magnesium}

\begin{figure}[h]
\centering
\includegraphics[width=0.68\linewidth]{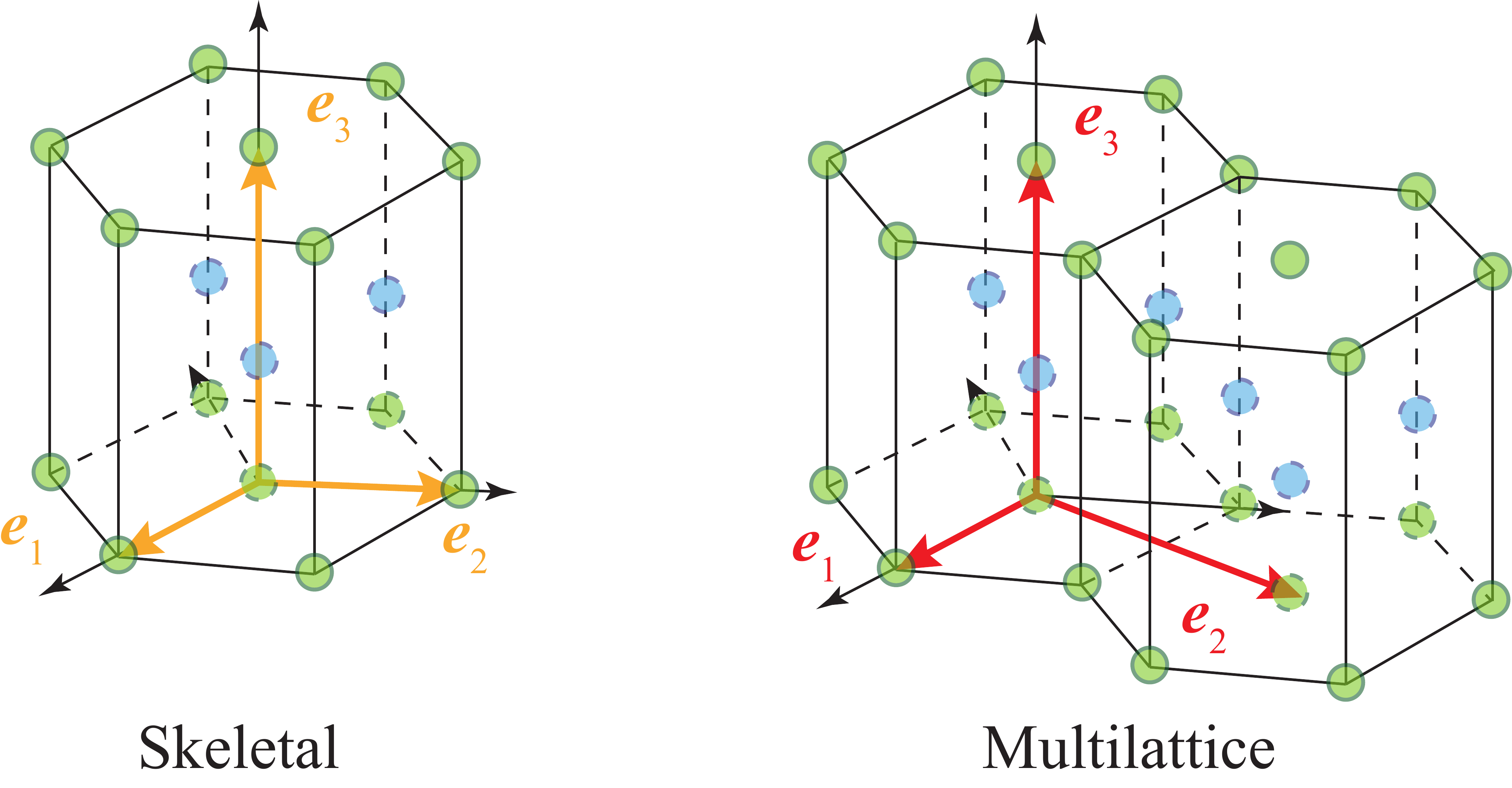} 
\caption{The unit and the super cell used in the calculation of the potential twinning modes of magnesium along with the choice of lattice vectors.}
\label{fig:hcp}
\end{figure}

We now apply the procedure described above to magnesium which has a hexagonal close packed structure with {lattice parameters $a = 3.196, c = 1.623$}.  This is not a Bravais lattice, and the smallest unit cell has two atoms.  We consider the basic unit cell and one super cell with four atoms in the unit cell shown in Figure \ref{fig:hcp}.  We also confine our search to matrices whose elements $|\mu_i^{\ j}| \le 4$.  This leads to almost $4 \times 10^8$ cases.  For each case, we check if (\ref{eq:DeformationGradient}) can be solved, and the two solutions if possible.  We find thousands of solutions even after collapsing those related by symmetry.

\begin{figure}[t]
\centering
\includegraphics[width=0.48\linewidth]{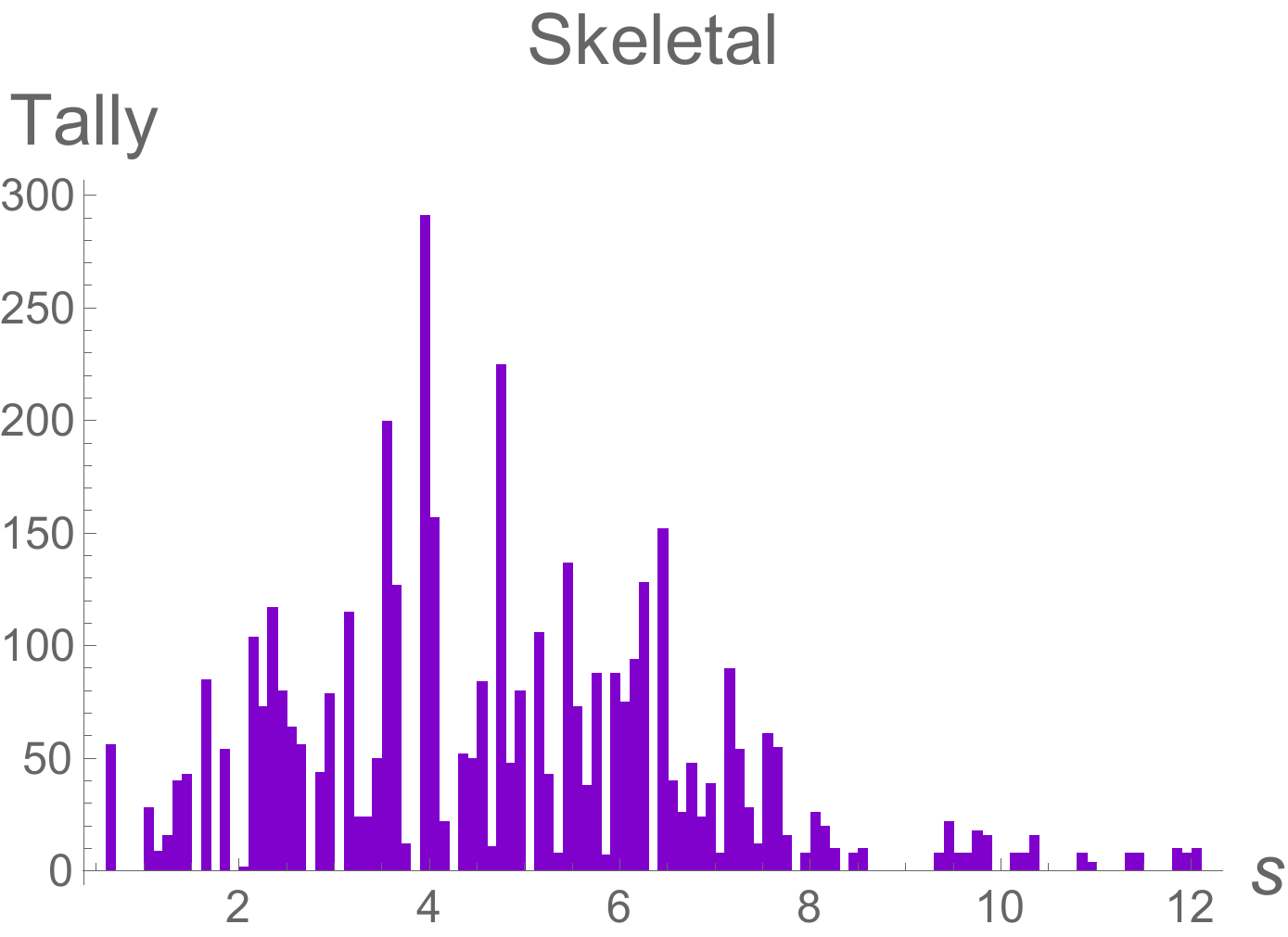} 
\includegraphics[width=0.48\linewidth]{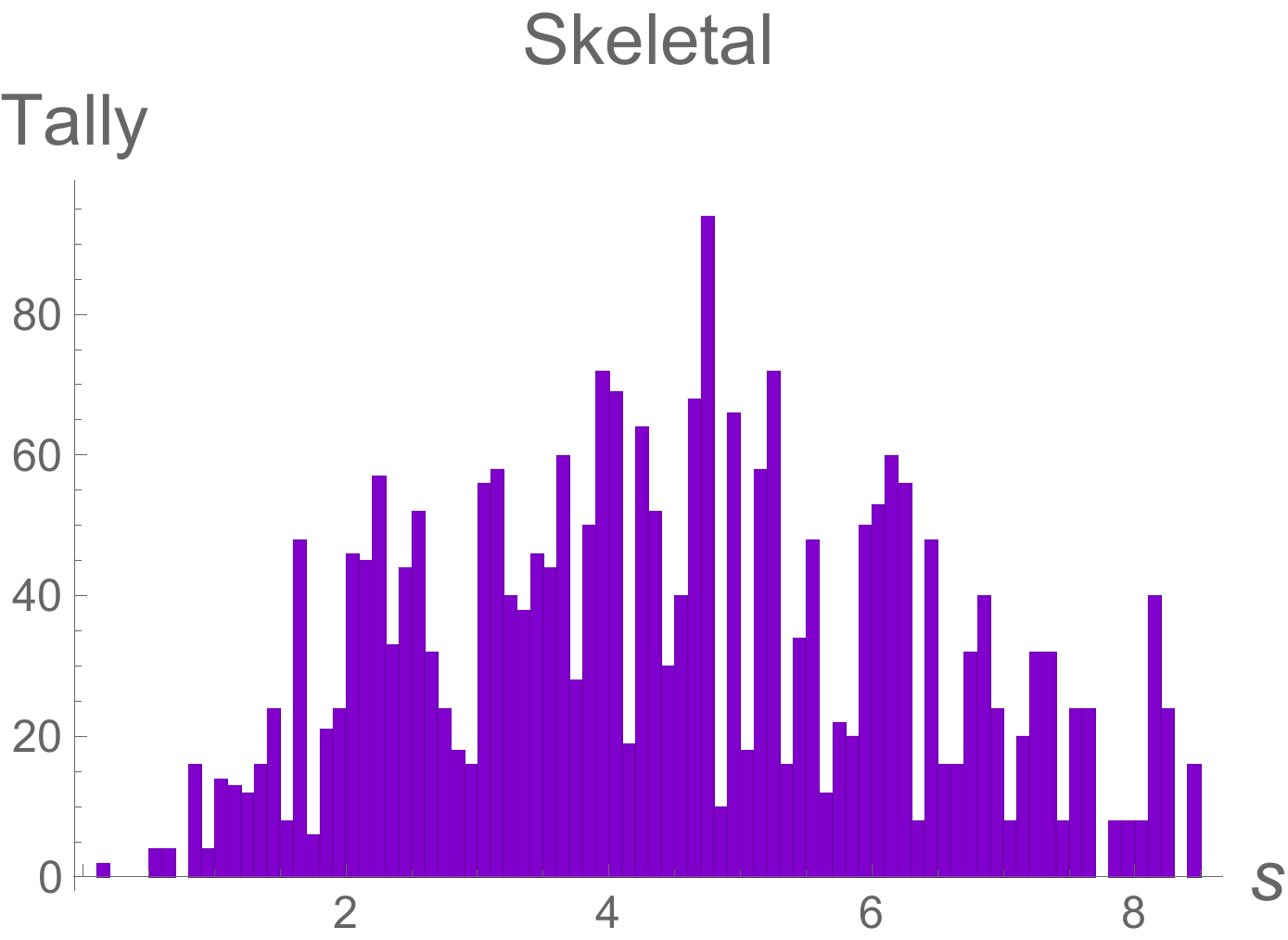}
\\
\includegraphics[width=0.48\linewidth]{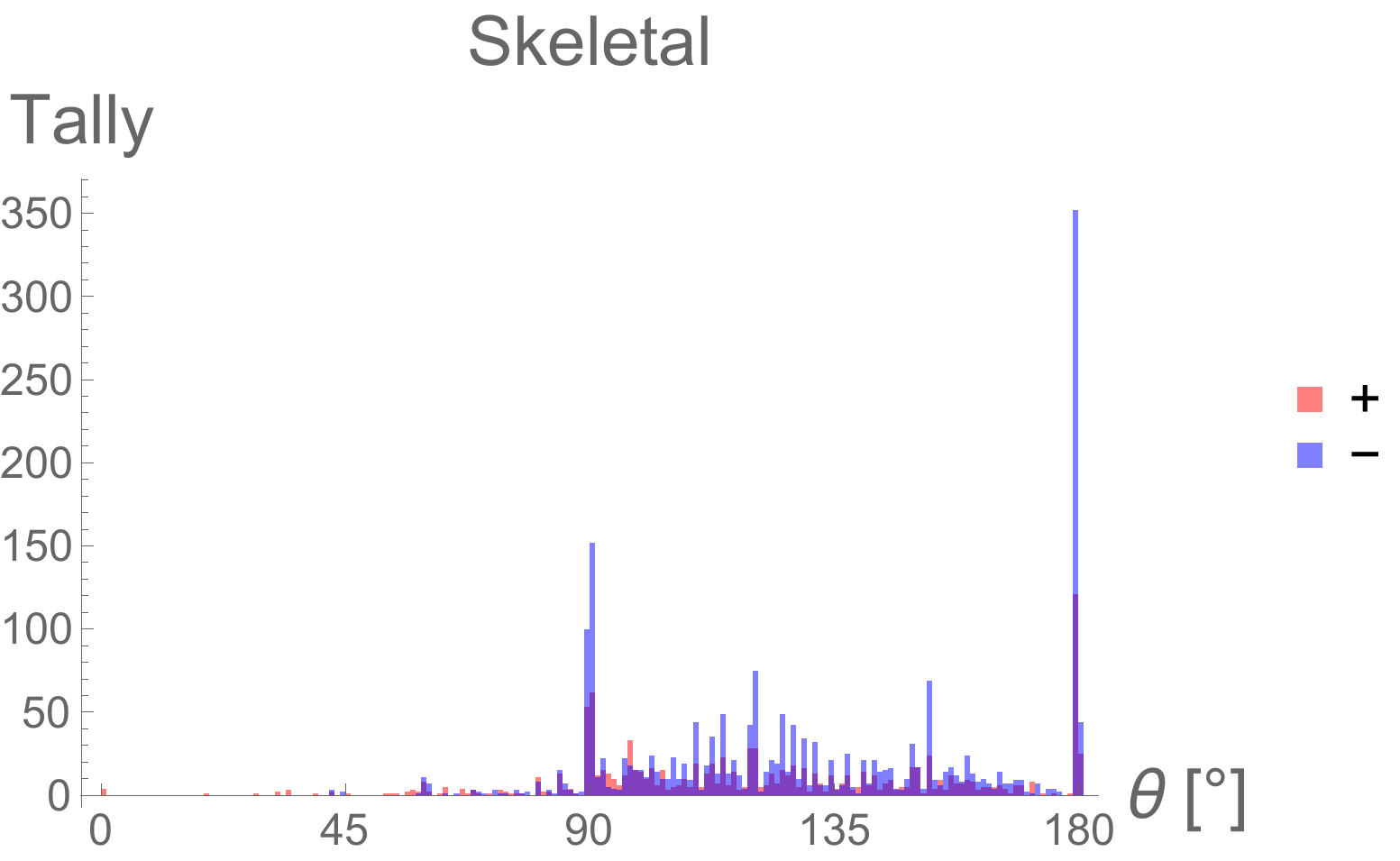}
\includegraphics[width=0.48\linewidth]{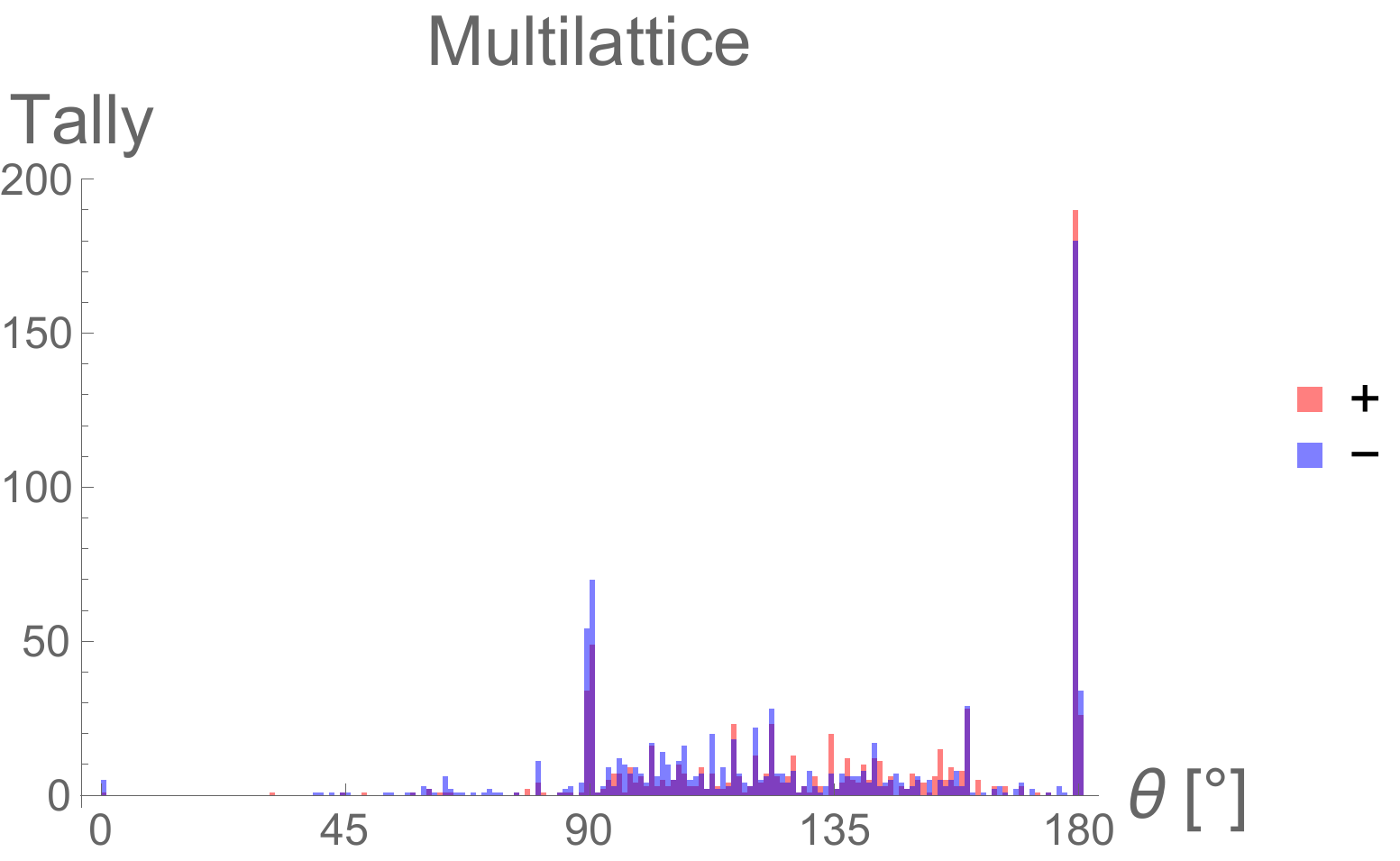}
\caption{Details of the many kinematically allowed twin modes of magnesium.  Top row: the histograms of shear magnitude.  Bottom row: Histogram of the angles of rotation.  The `+' and `$-$' refer to the conjugate twins. Since both solution branches of $\kappa=\pm 1$ do not affect the shear magnitude, we do not make the distinction when plotting the shear magnitudes.}
\label{fig:hist}
\end{figure}

Figure \ref{fig:hist} shows a histogram (after symmetry reduction) of the magnitude of shears and the angles of rotation $\theta$ of $\mathbf{Q}$.  A vast majority of modes have an un-physically large magnitude of shear.  The smallest magnitude of shear is 0.129 and this occurs for the well-known $\{10\bar{1}2\}\langle10\bar{1}1\rangle$ tension twin.  This is a compound twin and so its conjugate is itself.   However, notice that there are a large number of twinning modes whose magnitude of shear is smaller than or comparable to that (1.4919) of the well-known $\{10\bar{1}1\}\langle10\bar{1}2\rangle$ compression twin.  This is our first indication that other twin modes may be active in hexagonal close-packed materials like magnesium.

This proliferation of kinematically allowed twinning modes with comparable shears is very different to what we find in body centered cubic and face centered cubic systems \cite{SunThesis} where the mode with the smallest shear is vastly separated from all other modes.

As noted earlier, the framework above does not restrict the twin rotations to be two-fold.  We see a range of rotations, but there is a very large number of modes that happen to be two-fold ($\theta = 180^\circ$).  We also see additional peaks in the histogram at the crystallographic four fold ($\theta = 90^\circ$), three fold ($\theta = 120^\circ$) and six fold ($\theta = 60^\circ$) rotations.  We also note that conjugate modes may have different rotation angles.  Similarly, our framework does not restrict the search to have rational twin planes, and indeed a number of solutions do have irrational twin planes.

We have also noted earlier that twin modes can share the same twin plane and shear direction, but different shear magnitudes.  We observe this in our solutions.  For example, we find multiple modes with $\{10\bar{1}2\}\langle10\bar{1}1\rangle$.  The smallest one has a shear magnitude of $0.129$ as noted, but there is another one with shear magnitude $2.0042$.  We shall label twin modes that share the twin plane and shear direction as variants of the same mode.

\begin{figure}
\centering
\includegraphics[width=0.48\linewidth]{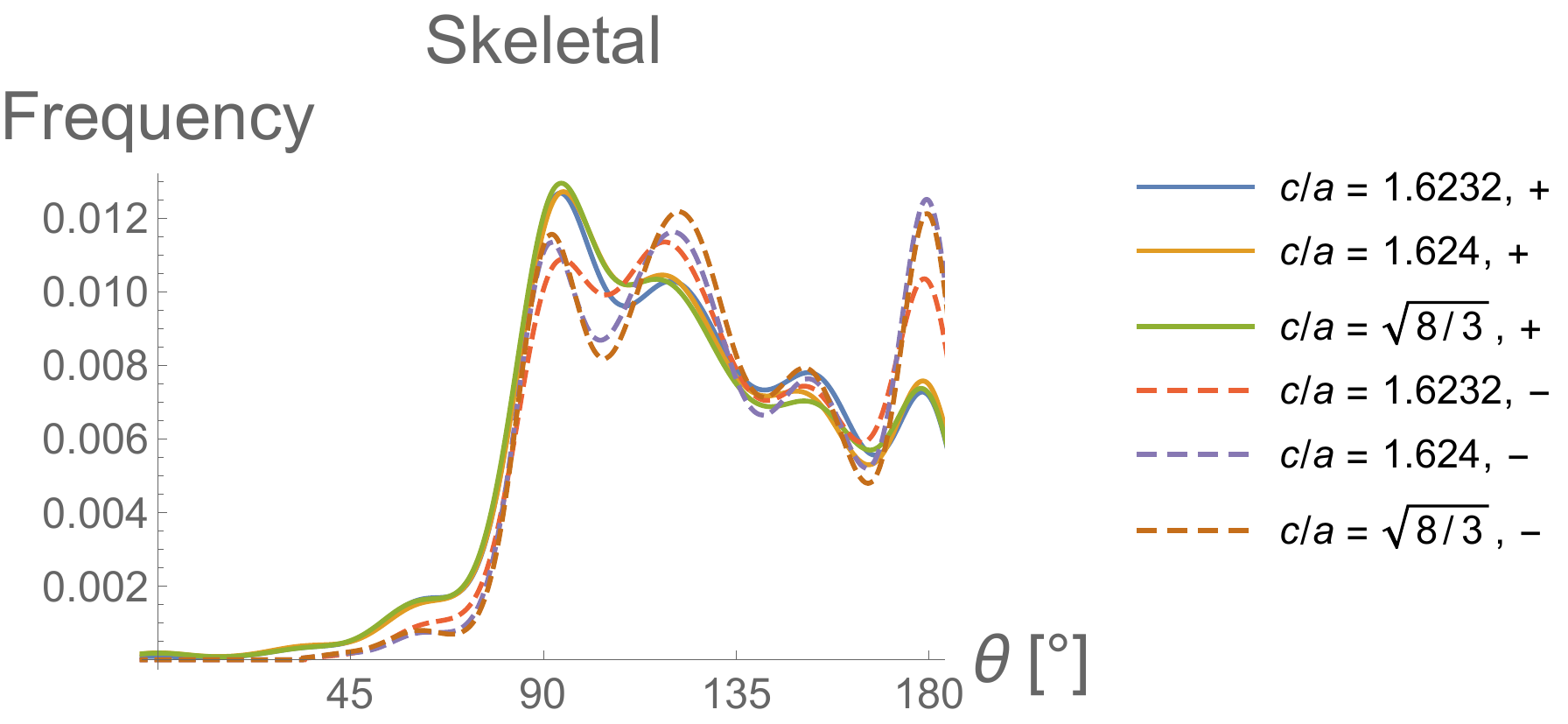}
\includegraphics[width=0.48\linewidth]{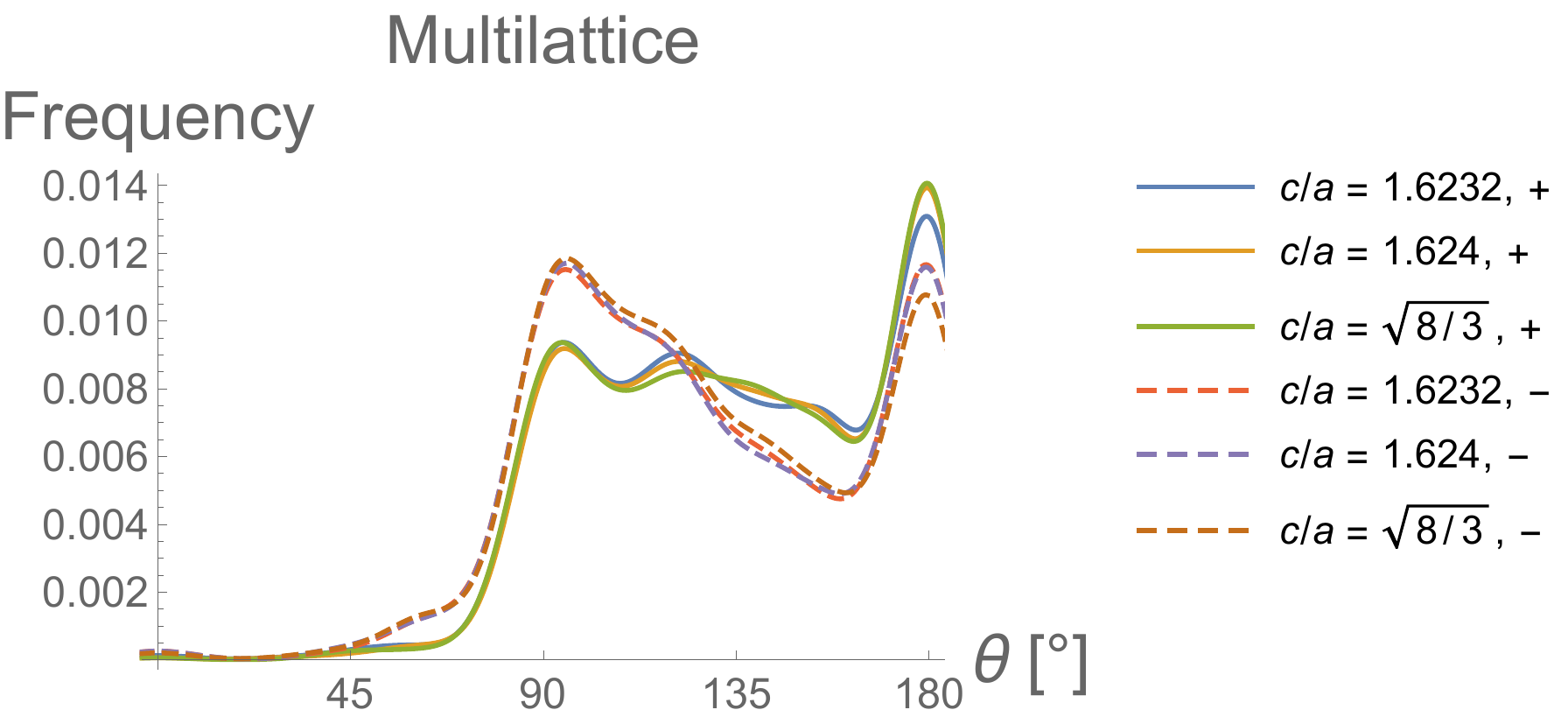}
\caption{Smoothed histogram of the angles of rotation for the twins found under the kinematic framework, using both the skeletal and multilattice constructions, for three different $c/a$ ratios.}
\label{fig:AnglesDistributionca}
\end{figure}

We conclude with a final observations.  The framework above is valid for all crystals, and in particular for all hexagonal close packed crystals.  However, the details depend on the $c/a$ ratio (since we can uniformly scale all calculations with $a$).  We demonstrate this in Figure \ref{fig:AnglesDistributionca}, where we show how the twinning angle changes with c/a ratio for the computed modes.

%%%%%%%%%%%%%%%%%%%%%%%%%%%%%%%%
%%%%%%%%%%%%%%%%%%%%%%%%%%%%%%%%
\section{Twin energetics}
\label{sec:ener}

The kinematic admissibility is a necessary condition for a twinning mode.  However, whether a material displays a twinning mode ultimately depends on the energy of the twinned configuration as well as the barrier to twinning.  We evaluate these in this section.  We first evaluate the twin boundary energy of the kinematically admissible modes identified in the previous section.  We then select a subset of the kinematically admissible modes with (relatively small) shear and twin boundary energy and evaluate the energetic barrier for the formation of these twins.

In order to evaluate the energy of the predicted twin modes, we use atomistic simulations based on molecular statics (MS). We primarily use the second nearest-neighbor Embedded Atom Method (2NN-MEAM) interatomic potential developed in Wu \emph{et al.} \cite{WuFrancisCurtin2015}. As the twin energy depends on the choice of the interatomic potential, we have also used other potentials, including the EAM potential of Sun \emph{et al.} \cite{SunEtAl2006},  MEAM potential of Kim \emph{et al.}\cite{KimEtAl2009} as well as an \emph{ab inito} electronic structure method MacroDFT \cite{Phanish2012,PongaBhattacharyaOrtiz2015}.  We find that, even though the fine results do depend on the choice of potential, the overall picture of a diverse energetic landscape remains approximately the same.  We discuss this further in the final section, and refer the reader to Sun \cite{SunThesis} for details.  For now we note that the parametrization we use yields material parameters shown in Table \ref{tab:TrainingHCP}.  

\begin{table}
\centering
\begin{tabular}{c c c }
\hline\\[-0.35cm]
{} & Modified MEAM \cite{WuFrancisCurtin2015} & Experiments\\
\hline\\[-0.35cm]
$a$ [\AA] & 3.196 & 3.209 \cite{WalkerMarezio1959}\\
$c/a$ &  1.623 & 1.624 \cite{WalkerMarezio1959}\\
$E_0^\text{pc}$ [eV/atom]  & -1.508 & -1.510 \cite{Kaxiras2003,Kittle2005,WachowiczKiejna2001}\\
$\gamma_{\{10\bar{1}2\}}$ [mJ/m$^2$] & 137 & 118 \cite{WangEtAl2010}\\
$C_{11}$ [GPa] & 67.7 & 63.5 \cite{SimmonsWang1971}\\
$C_{12}$ [GPa] & 24.7 & 25.9 \cite{SimmonsWang1971}\\
$C_{13}$ [GPa] & 18.7 & 21.7 \cite{SimmonsWang1971}\\
$C_{33}$ [GPa] & 68.9 & 66.5 \cite{SimmonsWang1971}\\
$C_{44}$ [GPa] & 17.9 & 18.4 \cite{SimmonsWang1971}\\
\hline
\end{tabular}
\caption{Materials details computed using modified MEAM. Comparisons are drawn to experimental values, except for $\gamma_{\{10\bar{1}2 \}}$, whose comparative value is obtained from \emph{ab-initio} simulations.}
\label{tab:TrainingHCP}
\end{table}

We conduct all our calculations using the software package LAMMPS \cite{LAMMPS}.

%%%%%%%%%%%%%%%%%%%%%%%%%%%%%%%%
\subsection{Twin boundary energy}
\begin{figure}
\centering
\includegraphics[width=0.5\linewidth]{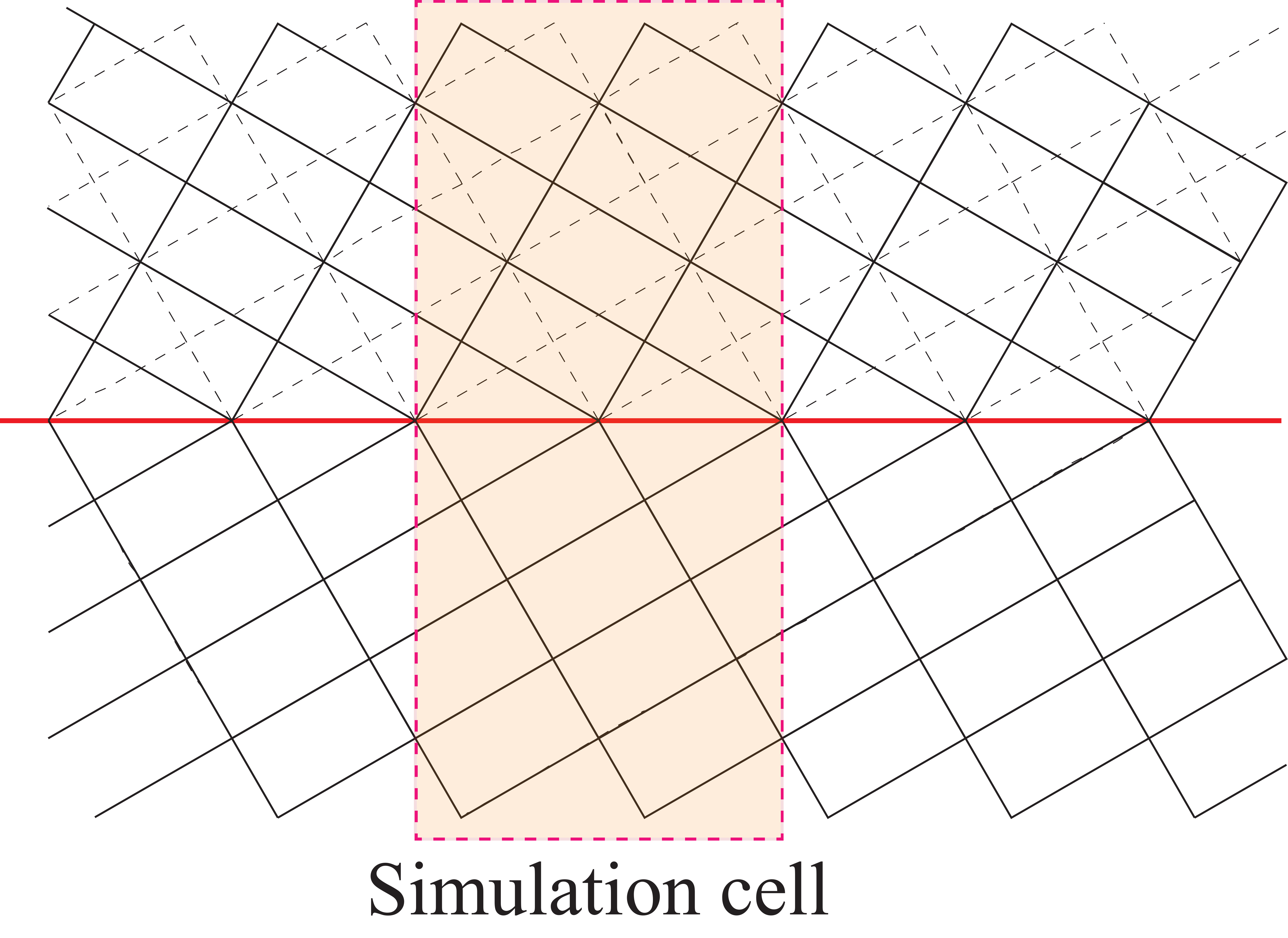}
\caption{Schematic illustration of the computational cell used to calculate the twin boundary energy.}
\label{fig:LatticeWithCell}
\end{figure}

\begin{figure}
\centering
\includegraphics[width=0.4\linewidth,angle=90]{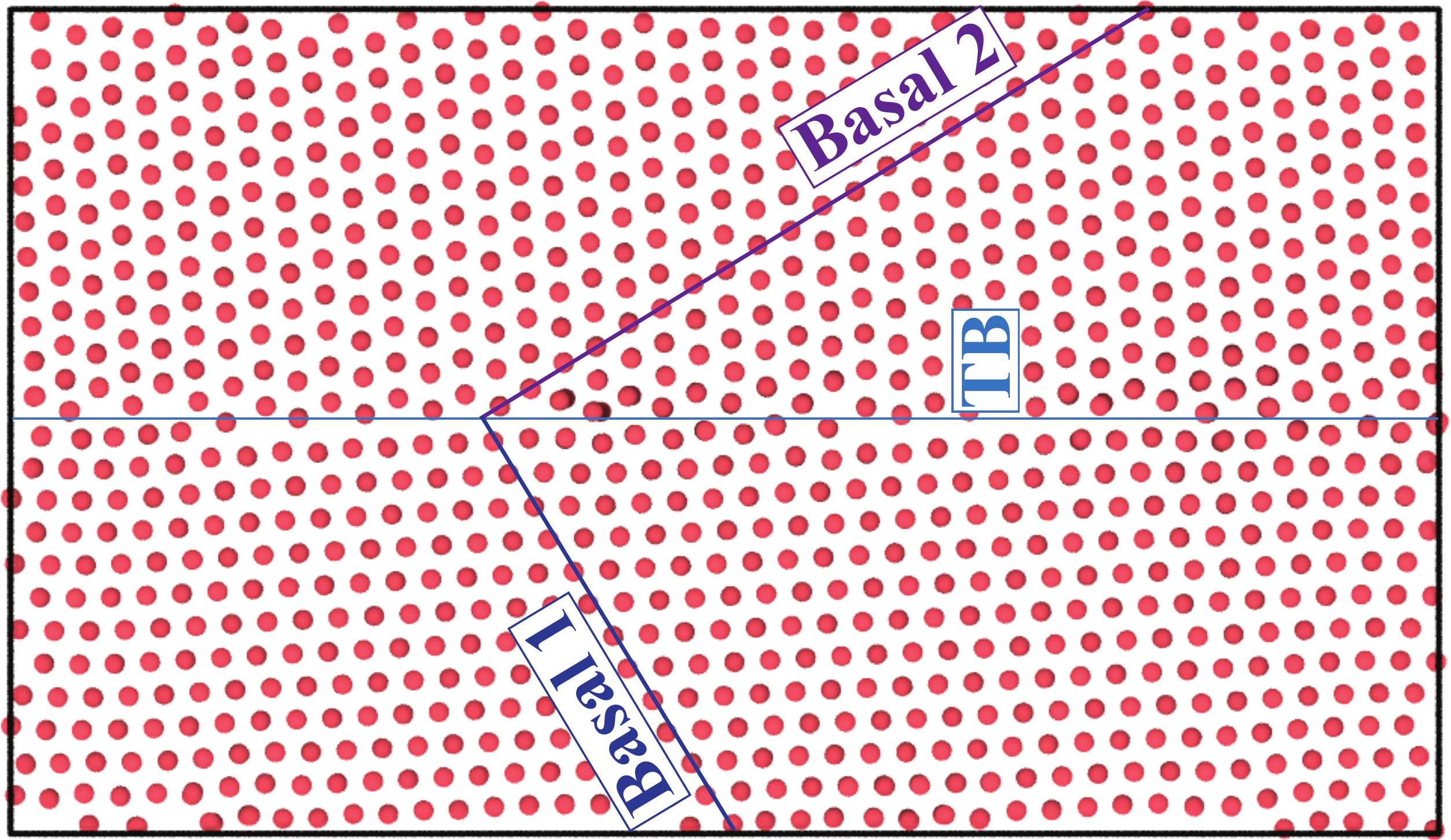}
\caption{Atomic configuration of a slice through the irrational twin system h with the (average) twin boundary and basal planes labeled.  Note that the twin boundary with irrational indices breaks into facets or steps.}
\label{fig:IrrationalTwin}
\end{figure}

We consider the computational cell shown schematically in Figure \ref{fig:LatticeWithCell} where one face is parallel to the twin plane and two faces are perpendicular to it.  We assume periodicity in the twin plane.  Note that this is exact when the twin plane is rational as there is a net of atoms while it is approximate when the twin plane is irrational with the approximation becoming more accurate with larger computational cells.  We start our simulations with the positions of the atoms to be those given by the idealized kinematic calculations, and relax the positions of the atoms or minimize the energy using  the Polak-Ribi\`{e}re conjugate gradient algorithm while constraining the top and the bottom atoms to translate only perpendicular to the twin wall.  Our typical computational cell contains about 6000 atoms, though we have verified that these results are accurate to cells containing as many as a million atoms.  We find that the atoms relax near the twin boundary and the displacements decay rapidly away from it. irrational twin boundaries or those with high index often break into facets.  An example is shown in Figure  \ref{fig:IrrationalTwin}.  We compute the twin boundary energy as
\begin{equation}\label{eq:GammaTwin}
\gamma^\text{tw}=\dfrac{E^\text{tw}(N)-E_0^\text{pc}N}{A_\text{interface}},
\end{equation}
where $E^\text{tw}(N)$ is the energy of the relaxed twin configuration with $n$ atoms, $E_0^\text{pc}$ is the binding energy per atom (i.e., the energy per atom of the untwinned specimen and $A_\text{interface}$ is the area of the twin plane within the computational cell.

\begin{figure}
\centering
\includegraphics[width=0.93\linewidth]{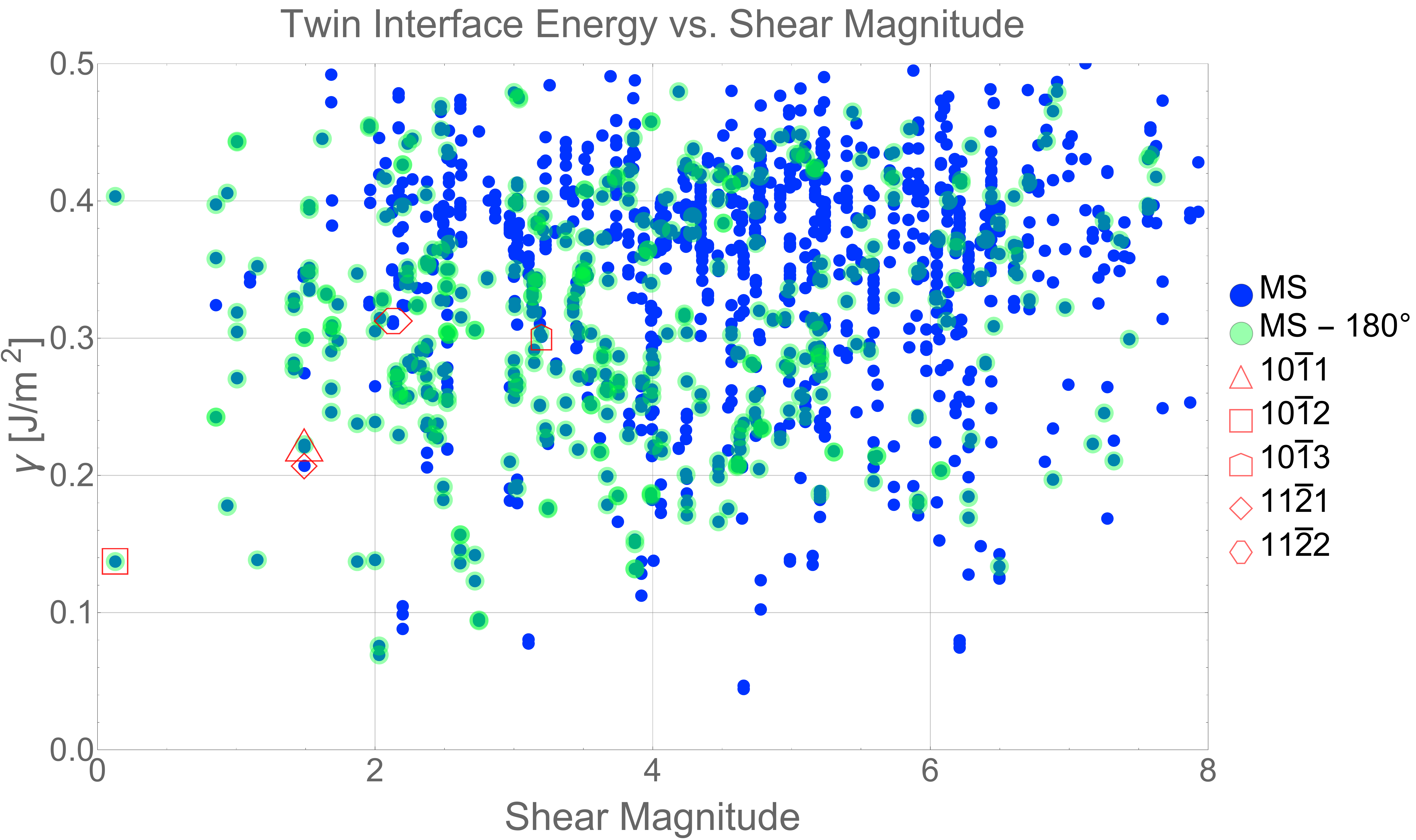}
\caption{Plot of all twin configurations with $\gamma^\text{tw}$ under 500 mJ/m$^2$ and calculated shear magnitude $s<8$ using the Modified MEAM potential by Wu \emph{et al.}. Twin solutions which satisfy two-fold rotations are marked by a teal outline. The two classical compression and tension twins are also highlighted. The modes reported in the literature are also highlighted.}
\label{fig:EnergyShearPlot}
\end{figure}

The computed results are shown in Figure \ref{fig:EnergyShearPlot} which shows the twin boundary energy $\gamma^\text{tw}$ and the shear magnitude for the various modes.  The tension and the compression twins are highlighted, as well as the other modes mentioned in the literature (please refer to the introduction of this work, Section \ref{sec:Intro}).  Twin modes corresponding to a two-fold rotation are also called out.

An important observation from this figure is that while the well-accepted tension twin stands out for having both low twin boundary energy and twinning shear, there is nothing remarkable about the other modes including the well-accepted compression twin.  In other words, a very large number of previously unstudied twinning modes have twin boundary energy and twinning shear that are comparable to the classical compression twins as well as the other twin modes mentioned in the literature.  We also notice that these relatively low energy - low shear modes predominantly involve two-fold rotations; however, non-two-fold rotations occur as well.

%%%%%%%%%%%%%%%%%%%%%%%%%%%%%%%%
\subsection{Energy barriers to twin formation}
\label{sec:energybarrier}

\begin{figure}[t]
\centering
\includegraphics[width=0.9\linewidth]{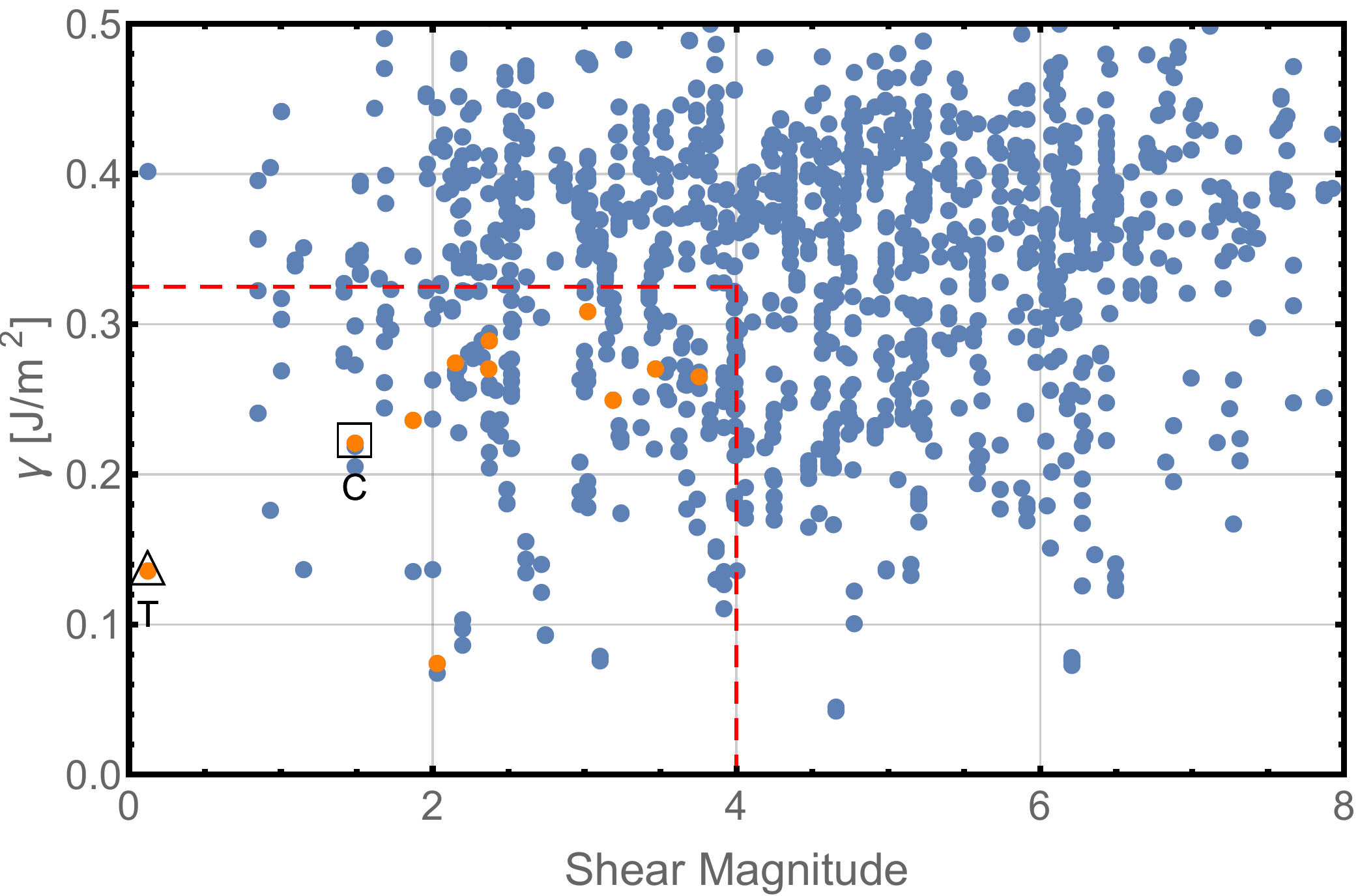}
\caption{Plot of all twin configurations with twin boundary energy $\gamma$ under 500 mJ/m$^2$ and calculated shear magnitude $s<8$ using the Modified MEAM potential by Wu \emph{et al.\ } Red dashed line denotes the boundary of points which were subjected to further nudged elastic band and stress calculation. Orange points denote twin systems which were detected to play a role in the yield surface of magnesium. The tension (T) and compression (C) twins are also highlighted.}
\label{fig:EnergyShearPlot2}
\end{figure}

A low twin interface energy and shear do not guarantee formation if the barriers to formation in the intermediate steps prove to be excessive.  Therefore we need to understand the energetic barriers to formation.  We focus on those twin modes which have twin boundary energy and shear comparable to the modes mentioned in the literature; we use $\gamma^\text{tw} \le$ 325 mJ/m$^2$ and $s \le 4$ identified by the dashed box in Figure \ref{fig:EnergyShearPlot2}. 

For each of the 229 modes in this region, we use the method of nudged elastic band to compute the energetic barrier between the perfect crystal and the twinned crystal.  The method of the nudged elastic band is described in detail in Henkelman and Jonsson \cite{HenkelmanJonsson2000}, Henkelmen \emph{et al.} \cite{HenkelmanUberuagaJonsson2000}, and Nakano \cite{Nakano2008}, and seeks to find the low energy path from one state to another.  The key idea is to find intermediate points such that the pathway satisfies necessary geometric conditions on the energy landscape.  

For each intermediate configuration, we compute the change in energy per atom from the perfect crystal reference state,
\begin{equation}\label{eq:ChangeEnergy}
\Delta E=\dfrac{E^\text{tw}(N)}{N}-E_0^\text{pc},
\end{equation}
as well as the virial stress
\begin{equation}\label{eq:virial}
\sigma_{v,ij}=\dfrac{1}{\Omega}\sum_{k\in\Omega}\left[-m^k(v_i^k-\bar{v}_i)(v_j^k-\bar{v}_j)+\dfrac{1}{2}\sum_{l\in\Omega}(x_i^l-x_i^k)f_j^{kl} \right],
\end{equation}
where $k$ and $l$ index atoms in the domain of interest, $\Omega$ is the volume of the domain of interest, $m^k$ is the mass of the $k^\text{th}$ atom, $v_i^k$ is the $i^\text{th}$ component of the velocity of the $k^\text{th}$ atom, $\bar{u}_j$ is the $j^\text{th}$ component of the average velocity of atoms in the volume of interest, $x_i^k$ is the $i^\text{th}$ component of the position of atom $k$, and $f_i^{kl}$ is the $i^\text{th}$ component of force applied on atom $k$ by atom $l$\footnote{As the calculations are based on molecular statics, the kinetic energy term is not a factor in the virial stress for these calculations.}. We find that many of the modes identified in Figure \ref{fig:EnergyShearPlot2} have energy barrier comparable to the tension and compression twins.  Consequently, we investigate them further in the subsequent sections by studying the kinetic rate constants and the yield surface. Relevant information of these twins modes is shown in Table \ref{tab:TwinInfo}.

An interesting question in the study of magnesium and other hcp materials is whether the deformation modes change by the the application of pressure.  Therefore, we compute the energy barrier not only for specimens corresponding to the the relaxed lattice parameters but also for specimens subjected to equi-triaxial compression and tension up to 5\%.

%%%%%%%%%%%%%%%%%%%%%%%%%%%%%%%%
%%%%%%%%%%%%%%%%%%%%%%%%%%%%%%%%
\section{Kinetic rates}
\label{sec:kinetic}

\begin{figure}
\centering
\includegraphics[width=\linewidth]{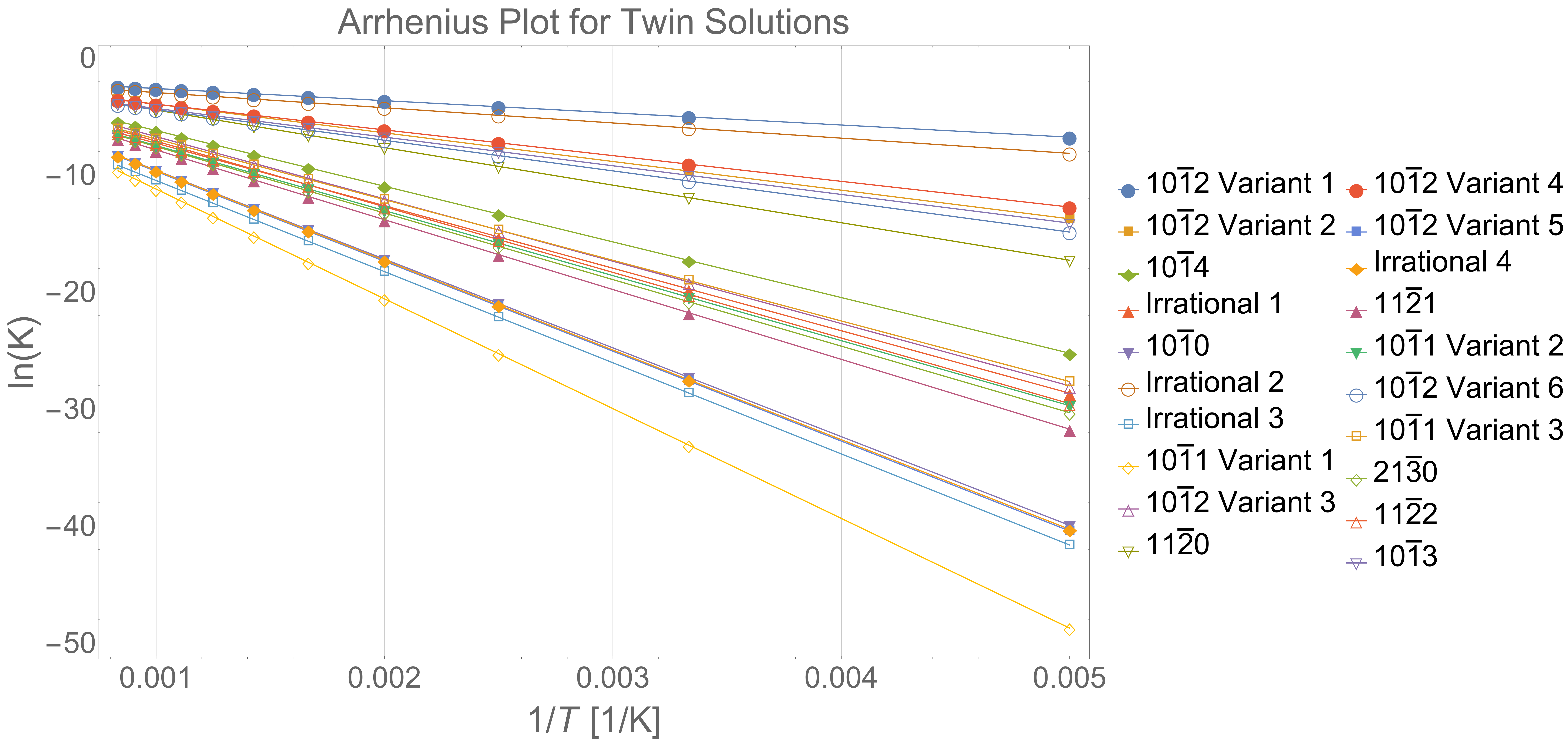}
\caption{Arrhenius plot of the various twin solutions with $s\le 2$ and $\gamma\le 300$ mJ/m$^2$ from 200-1200 K. {}}
\label{fig:NEBArrhenius}
\end{figure}

In the previous sections, we have found a large number of kinematically admissible states with comparable energetics.  This suggests that the deformation process is dominated by a number of competing modes, and therefore it is important to know the kinetics of these modes.  To gain an insight into the potential kinetics, we  assume that this is a thermally activated process and therefore the reaction rate is given by the Arrhenius relation:
\begin{equation}\label{eq:ArrheniusRelation}
K=\nu_0\mathrm{e}^{-\Delta E/(k_BT)},
\end{equation}
where $k_B$ is the Boltzmann constant,  $T$  the absolute temperature and $\nu_0$. We can compute the latter following \cite{Weiner2002} to be
\begin{equation}\label{eq:AttemptFrequency}
\nu_0=\dfrac{1}{2\pi}\dfrac{\prod_{i=1}^n \omega_{i,S}}{\prod_{i'=2}^n\omega_{i',U}},
\end{equation}
where $\omega_{i,S}$ is the $i^\text{th}$ natural frequency (eigenvalue) in the stable (perfect crystal) state, and $\omega_{i,U}$ representing the $i^\text{th}$ eigenvalue in the metastable state\footnote{In the denominator, the sum excludes the unstable eigenvalue, hence the different lower bound of summation.}.  In our setting, this is given by the configuration with the highest energy barrier in our nudged elastic band calculation. 

\begin{figure}
\centering
\subfigure[]{\includegraphics[width=\linewidth]{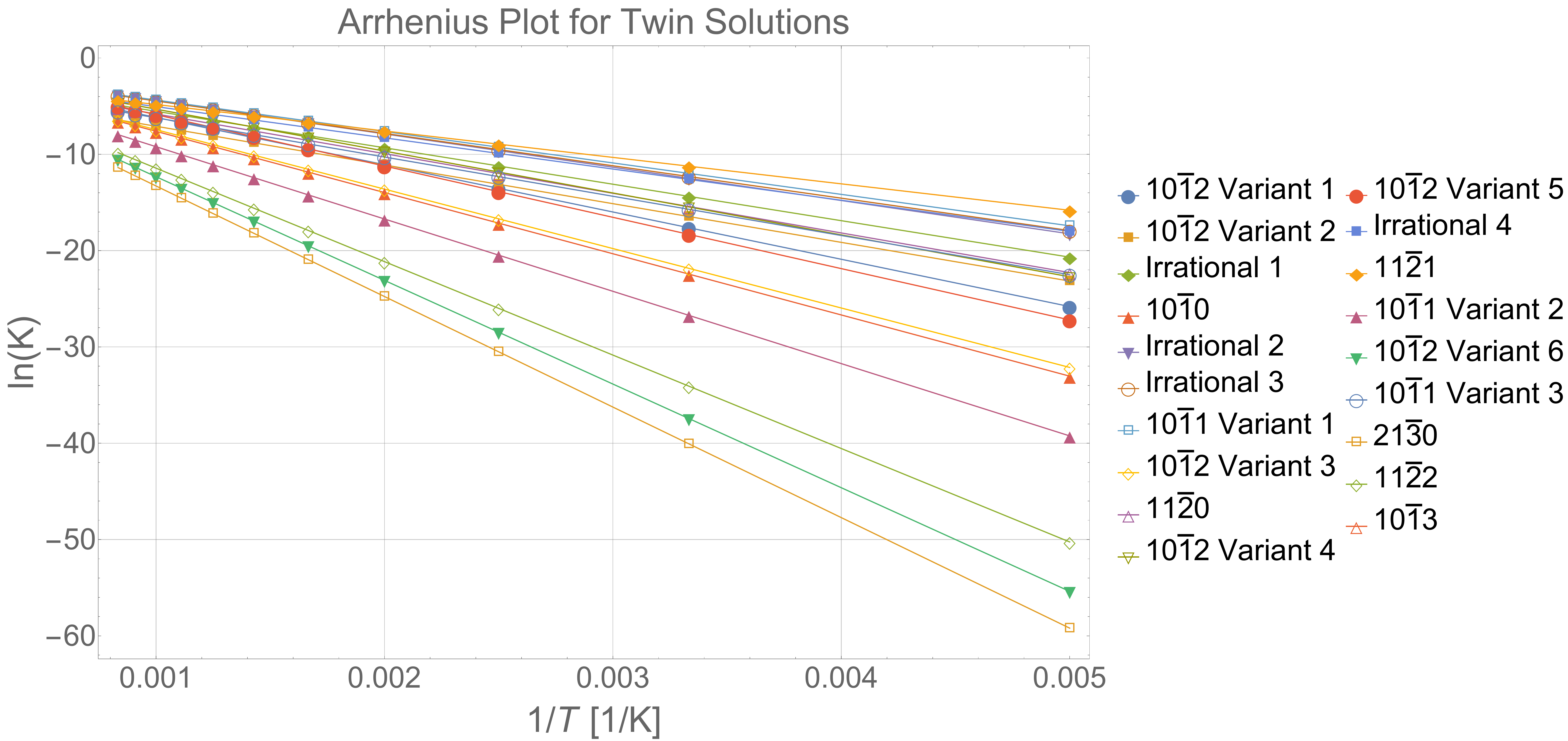}}
\subfigure[]{\includegraphics[width=\linewidth]{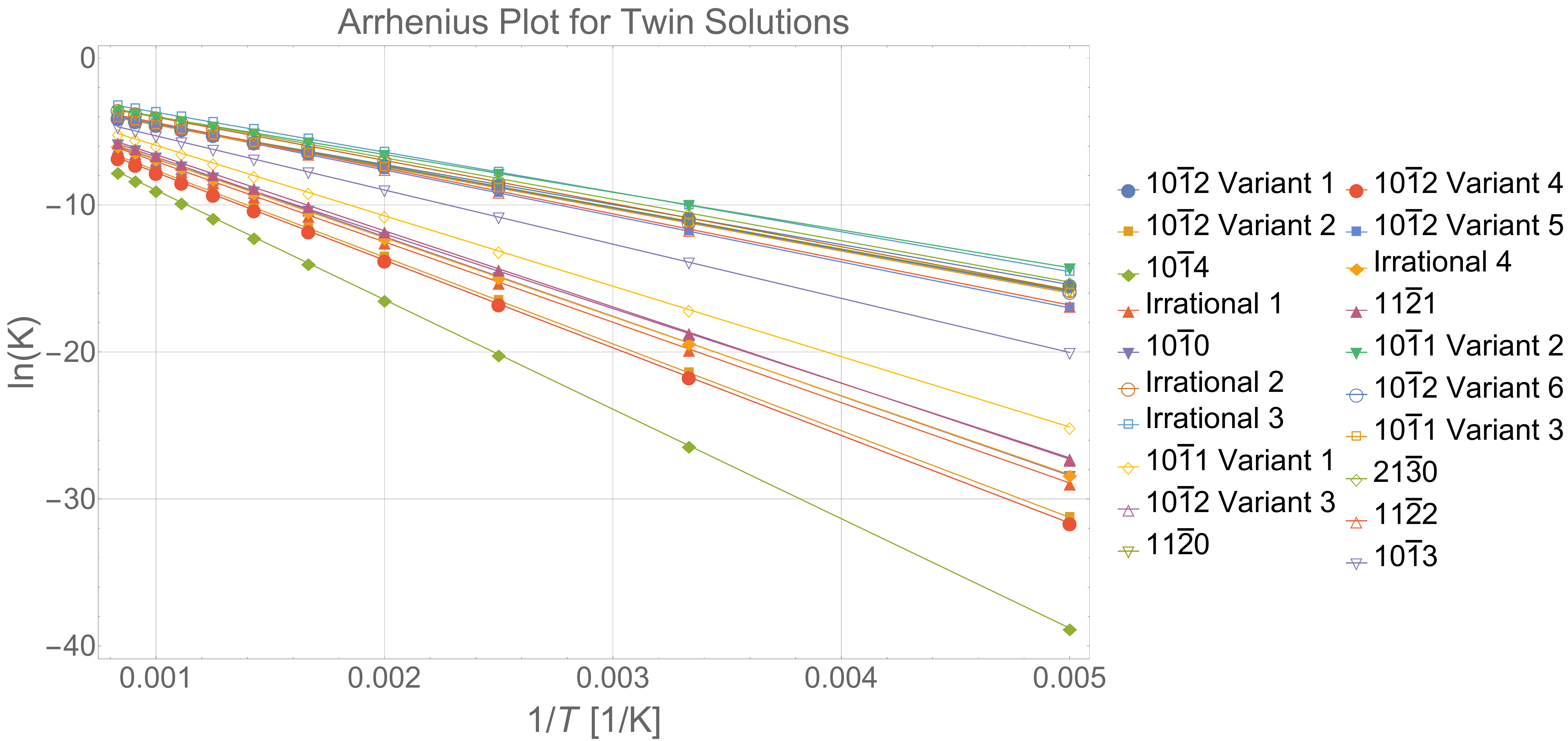}}
\caption{Arrhenius plot for same twins as Figure \ref{fig:NEBArrhenius} with sample under load, (a) displaying 3\% compression and (b) displaying 3\% tension.}
\label{fig:NEBArrheniusLoad}
\end{figure}

The reaction rates of twenty representative modes are plotted as a function of (inverse) temperature in Figure \ref{fig:NEBArrhenius}. {These twenty modes had $s\le 2$ and $\gamma\le 300$ mJ/m$^2$, encompassing the classical tension and compression twins for good comparison.}  Interestingly, notice that the classical tension mode has the highest rate.  However, a mode with an irrational twin plane is a close second.  The classical compression twin has the smallest rate amongst all the twenty twin modes considered.  

Figure \ref{fig:NEBArrheniusLoad} shows how the reaction rates change on the application of 3\% equi-triaxial compression and tension.  We see that the relative rates change with strain.  We also see that some of the curves cross each other.  Together, these show that twin activity and rates can depend on both pressure and temperature.

%%%%%%%%%%%%%%%%%%%%%%%%%%%%%%%%
%%%%%%%%%%%%%%%%%%%%%%%%%%%%%%%%
\section{Yield surface}
\label{sec:yield}

We now use the results of the previous section to compute the elastic domain or the yield surface.  As we apply stress to a crystal, we expect the material to remain elastic for some loads, and for certain deformation modes to activate as the stress reaches a critical value.  

When a stress is applied to a crystal, the driving force on the $i$th twin system depends on the resolved shear stress
\begin{equation}
\sigma^i = \mathbf{\hat{s}}_i\cdot \boldsymbol{\sigma}\mathbf{\hat{n}}_i,
\end{equation}
where ${\mathbf s}_i$ and $\hat{\mathbf n}_i$ are the twinning shear and twin plane associated with the $i$th twin system and $\boldsymbol{\sigma}$ is the Cauchy stress.  Since the Cauchy stress cannot be defined in atomic systems, we instead use the virial stress\footnote{We have computed and ensured that the cell sizes are sufficiently large enough that the computed stress for the system has converged.}. We expect the $i$th twinning system to be activated when this resolved shear stress $\sigma_i$ reaches a critical value $\sigma^i_c$.  We can obtain this critical resolved shear stress from our calculation in Section \ref{sec:energybarrier} of the virial stress along the transition path as the maximum value of the resolved virial stress:
\begin{equation}\label{eq:CRSS}
\sigma_c^i= \max \ \mathbf{\hat{s}}_i\cdot \boldsymbol{\sigma}_v\mathbf{\hat{n}}_i,
\end{equation}
where the maximum is taken along the transition path and ${\sigma}_v$ is the computed virial stress at any point on this path.  We calculate the critical resolved shear stress for all 229 potential twinning modes identified earlier.  For example, we compute a value of 20.8 MPa for the classical tension twin and this agrees well with the value of 18 MPa based on experiments \cite{StaroselskyAnand2003}.  This provides us confidence in our calculations.

\begin{table}
\centering
\label{tab:SlipAnd1012}
\begin{tabular}{c c c c c c}
\hline\\[-0.35cm]
System & Magnitude [MPa]\\
\hline\\[-0.35cm]
Basal & 0.52\cite{ConradRobertson1958}-0.55\cite{StaroselskyAnand2003}\\
Prismatic & 39.2\cite{ReedHillRobertson1957b}\\
Pyramidal & 105\cite{StaroselskyAnand2003}\\
\hline
\end{tabular}
\label{tab:slip}
\caption{Slip systems and critical resolved shear stress along these systems in magnesium.}
\end{table}

We are now in a position to define the elastic domain as the set of stress for which the material remains elastic since no system is activated:
\begin{equation}\label{eq:ElasticDomain}
\mathcal{Y}=\{\boldsymbol{\sigma}:\boldsymbol{\sigma=\sigma}^T,\operatorname{tr}\boldsymbol{\sigma}=0,\mathbf{\hat{s}}_i\cdot \boldsymbol{\sigma}\mathbf{\hat{n}}_i<\sigma_c^i\text{ for }i=1,\ldots,N  \}.
\end{equation}
The yield surface is the boundary $\partial \mathcal{Y}$ of the elastic domain and is the stress at which at least one system becomes active.   We make a few observations before we compute this set.   First, slip modes also contribute to the deformation of magnesium, and therefore, we have to append slip modes to the twinning modes under consideration.  Therefore, we add the three modes listed in Table \ref{tab:slip}.  Second, recall that each system can have multiple symmetry related manifestations.  We have to consider all manifestations in this calculation.  

Finally, it is possible that one system completely obscures another.  In other words, it is possible that not all systems participate in the definition of the elastic domain $\mathcal{Y}$.  Indeed, define 
\begin{equation}
{\boldsymbol G}_i = \frac{\mathbf{s}_i \otimes \hat{\mathbf n}_i}{\sigma_c^i}
\end{equation}
for the $i$th system.  Suppose ${\boldsymbol G}_i$ can be written as a convex combination of the corresponding tensors of a number of other systems: i.e.,
\begin{equation}
{\boldsymbol G}_i = \sum_{j \in I} \lambda_j {\boldsymbol G}_j
\end{equation}
for some set of systems $I$ that does not include $i$.
Then notice that ${\boldsymbol \sigma} \cdot {\boldsymbol G}_j < 1$ for all $j \in I$ implies that 
${\boldsymbol \sigma} \cdot {\boldsymbol G}_i < 1$.  It follows that if for some ${\boldsymbol \sigma}$,
$\mathbf{s}_j \cdot \boldsymbol{\sigma}\mathbf{\hat{n}}_j < \sigma_c^j$ for all $j \in I$, then $\mathbf{s}_i \cdot \boldsymbol{\sigma}\mathbf{\hat{n}}_i < \sigma_c^j$.  In other words, we do have to consider the $i$th system in our calculation of the yield surface.  We conclude that we only need to consider those systems that form the extreme points of the convex hull of the set of ${\boldsymbol G}_i$ for all $i$.  Further, the extreme points correspond to the active systems.

\begin{table}
\centering
\small
\begin{tabular}{c c c c c c c c c}
\hline\\[-0.35cm]
$i$ & Twin $\{K_1\} \langle \eta_1 \rangle$ & $s$ & $\theta$ & $\gamma$ [mJ/m$^2$] & $\Delta E$ [eV] & $\nu_0$ & $\sigma_Y$ [MPa] \\
\hline\\[-0.35cm]
a & $\{10\bar{1}2 \}\langle10\bar{1}1\rangle$ & 0.1299 & 180 & 137.0 & 0.0896 & 0.2086 & 20.8 \\
b & $\{10\bar{1}1 \}\langle10\bar{1}2\rangle$ & 1.4919 & 180 & 222.0 & 0.8084 & 0.1619 & 185.8 \\
c & $\{21\bar{3}0 \}\langle10\bar{1}0\rangle$ & 1.7321 & 180 & 297.4 & 0.4461 & 0.1689 & 125.9 \\
d & $\{11\bar{2}0\}\langle0001\rangle$ & 1.8743 & 180 & 242.2 & 0.3221 & 0.2058 & 188.9 \\
e & $\{21\bar{3}2 \}\langle10\bar{1}0\rangle$ & 2.0343 & 180 & 68.8 & 0.3001 & 0.2024 & 143.9 \\
f & $\{10\bar{1}3 \}_\text{I}\langle50\bar{5}4\rangle_\text{I}$ & 2.1568 & 180 & 275.4 & 0.6517 & 0.2372 & 63.6 \\
g & $\{21\bar{3}0 \}\langle10\bar{1}1\rangle$ & 2.3738 & 180 & 271.8  & 0.3507 & 0.2103 & 186.5 \\
h & $\{15,8,\overline{23},1 \}_\text{I}\langle10\bar{1}1\rangle$ & 2.3773 & 159.3 & 290.1 & 0.6900 & 0.1226 & 41.8 \\
i & $\{31,1,\overline{32},\overline{29} \}_\text{I}\langle10\bar{1}1\rangle$ & 3.0245 & 159.3 & 309.6 & 0.8424 & 0.1823 & 355.2 \\
j & $\{11\bar{2}1 \}_\text{I}\langle10\bar{1}1\rangle$ & 3.1921 & 159.3 & 250.6 & 0.4361 & 0.1573 & 211.6 \\
k & $\{30\bar{3}4 \}_\text{I}\langle10\bar{1}3 \rangle_\text{I}$ & 3.4703 & 180 & 271.4 & 0.2401 & 0.2717 & 11.9 \\
l & $\{10\bar{1}2 \}_\text{I}\langle40\bar{4}5\rangle_\text{I}$ & 3.7584 & 180 & 266.2 & 0.6210 & 0.1412 & 11.8 \\
\hline
\end{tabular}
\caption{Details of the twin systems which were found to affect the yield surface of magnesium. The first column is an arbitrary label. Mode a is the classical tension twin while mode b is the classical compression twin. The subscript I denotes an irrational index which has been rounded to a nearby integer.}
\label{tab:TwinInfo}
\end{table}

We use this procedure to identify the active systems and the elastic domain (equivalently yield surface).  We find that as many as {\it twelve twinning systems are active} and these are listed in Table \ref{tab:TwinInfo}.  We see the classical tension twin (mode a) and the classical compression twin (mode b).  However we see ten other systems.

\begin{figure}
\centering
\subfigure[]{\includegraphics[width=0.35\linewidth]{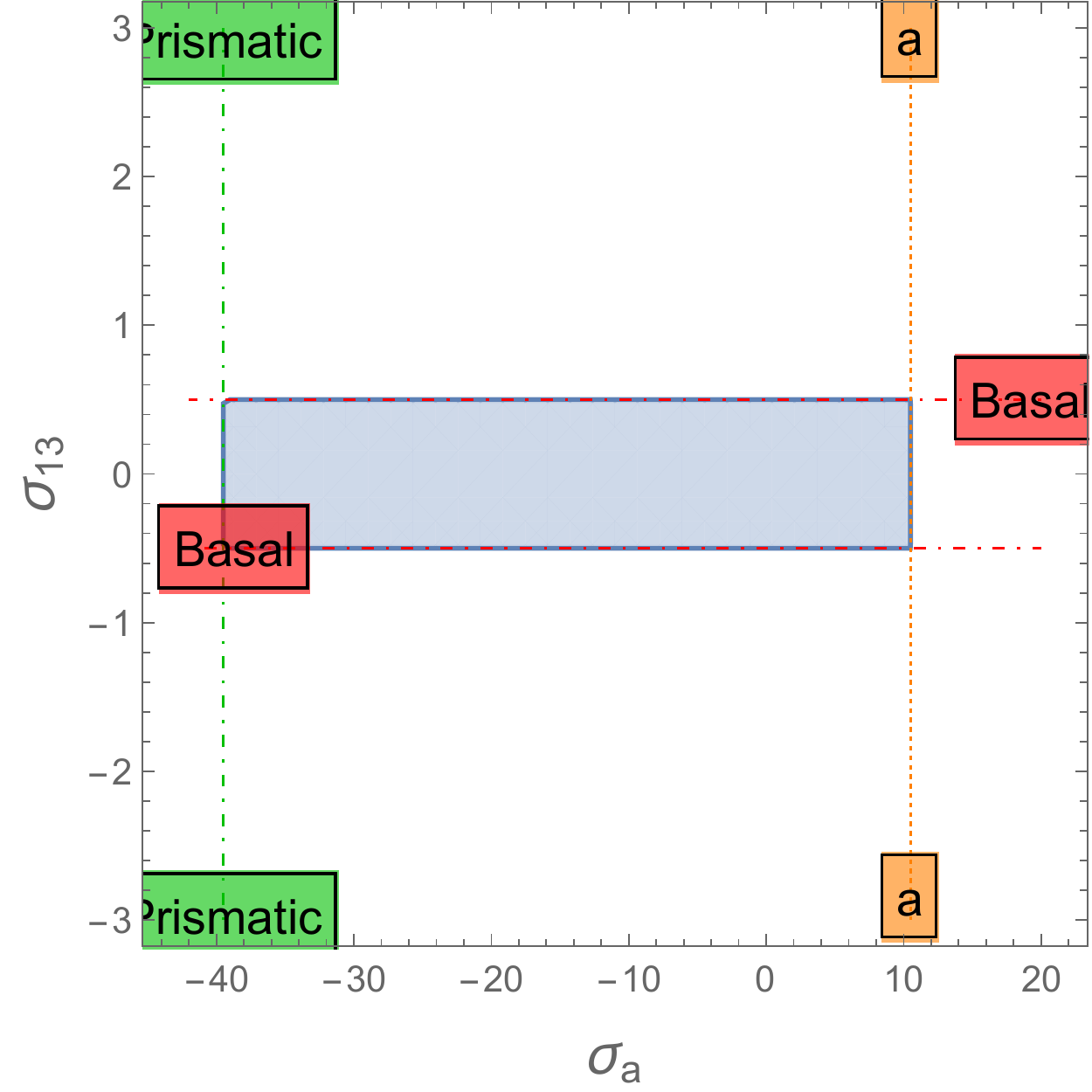}}
\subfigure[]{\includegraphics[width=0.35\linewidth]{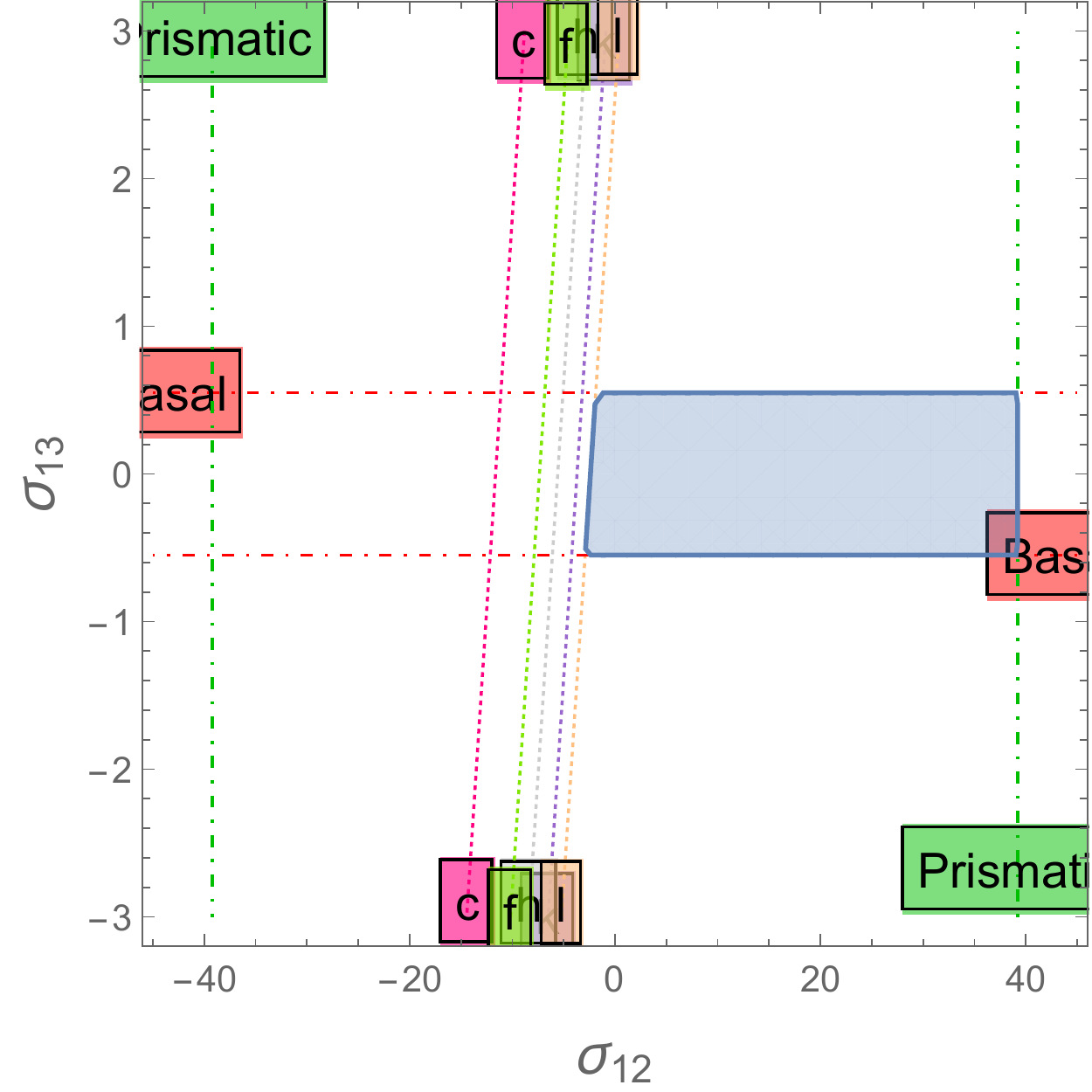}}
\subfigure[]{\includegraphics[width=0.35\linewidth]{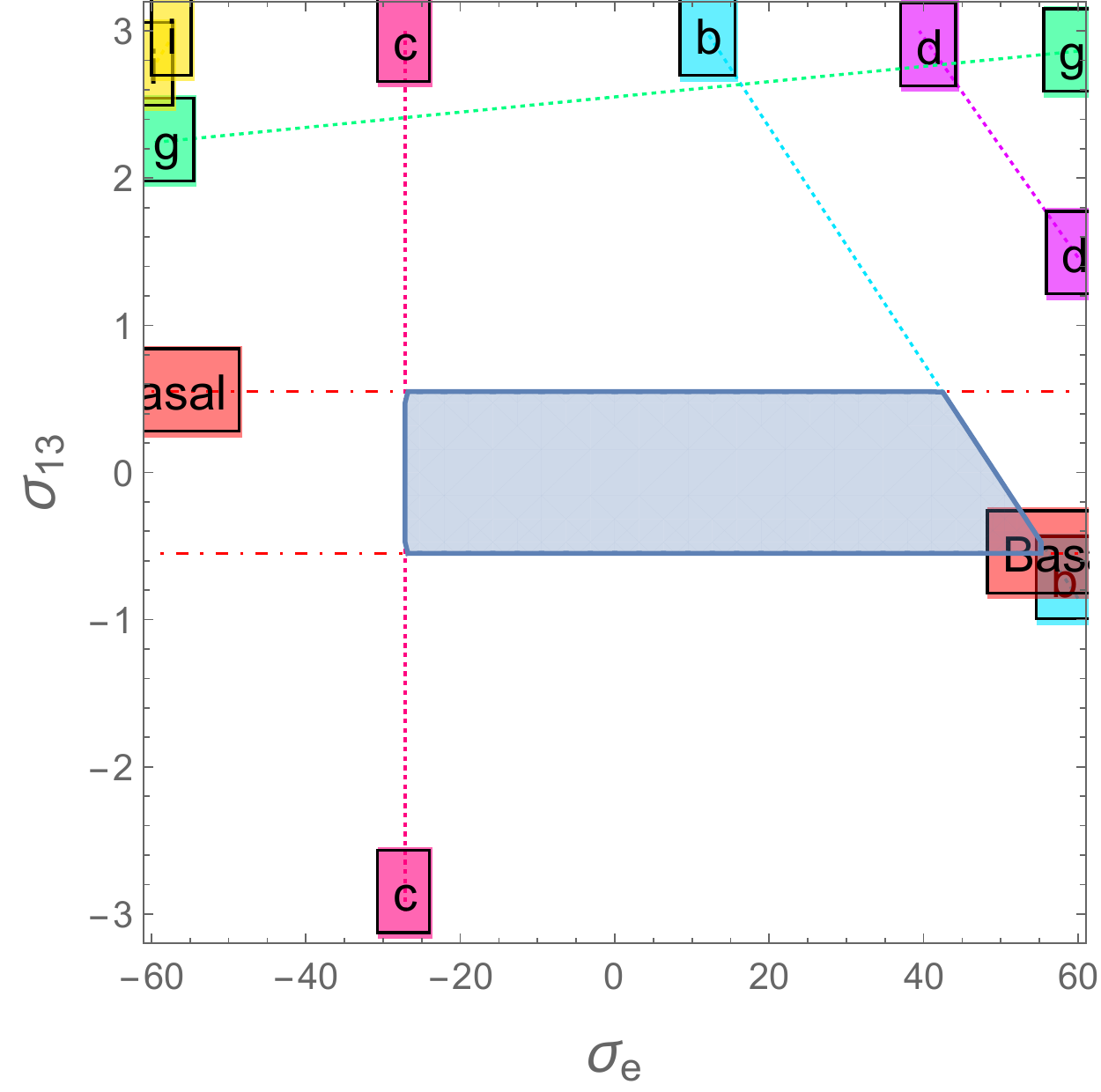}}
\subfigure[]{\includegraphics[width=0.35\linewidth]{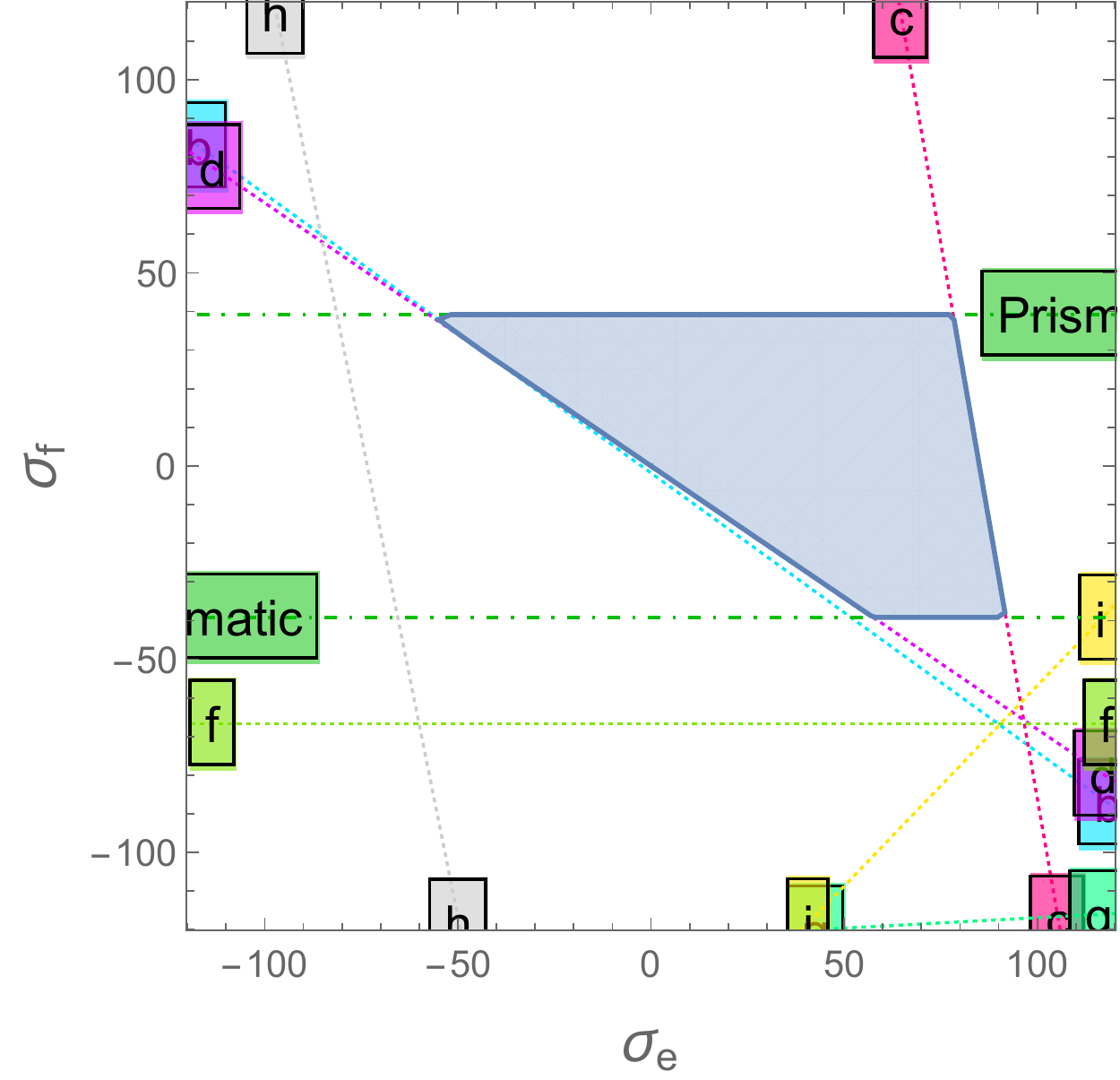}}
\subfigure[]{\includegraphics[width=0.35\linewidth]{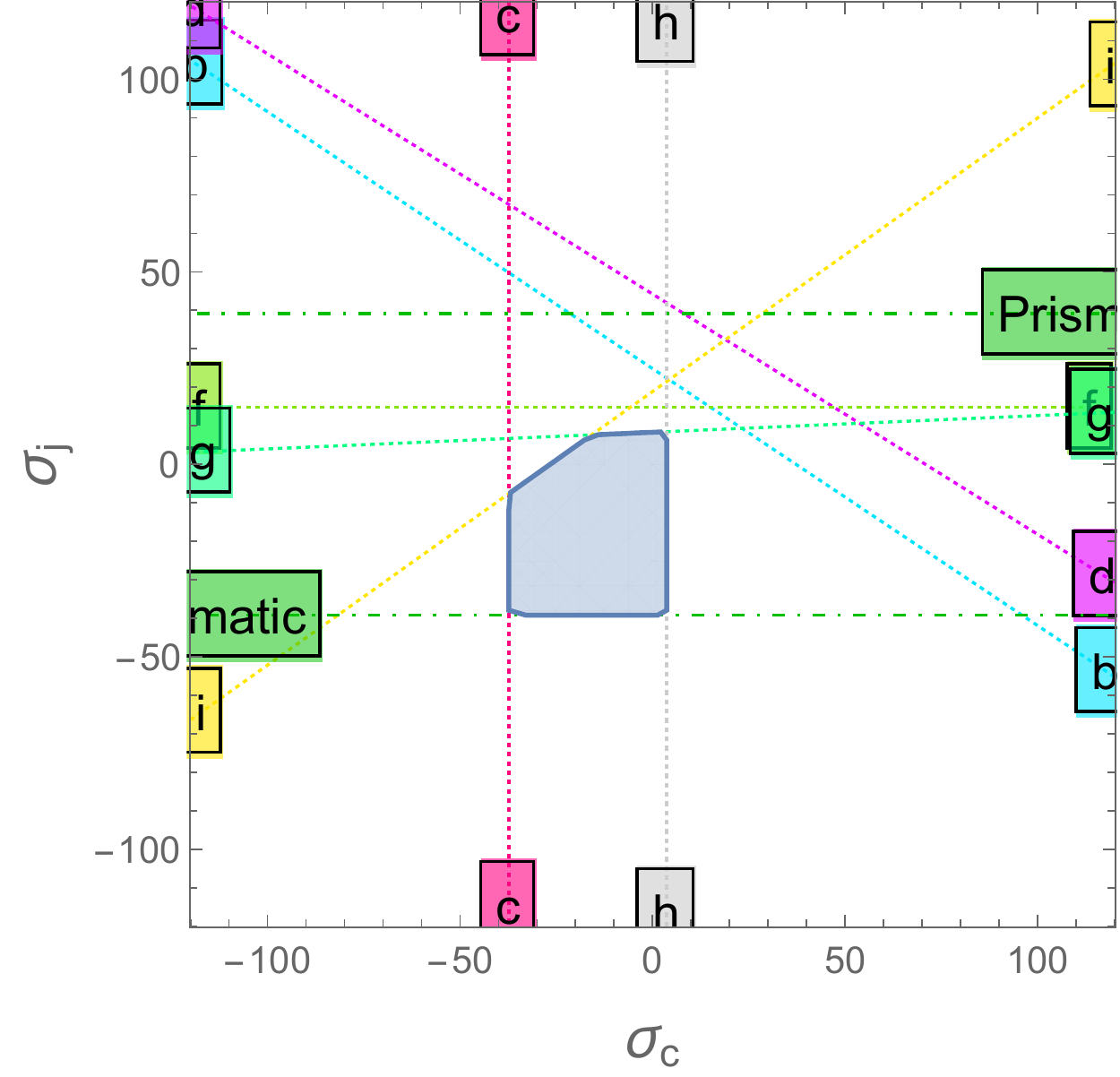}}
\subfigure[]{\includegraphics[width=0.35\linewidth]{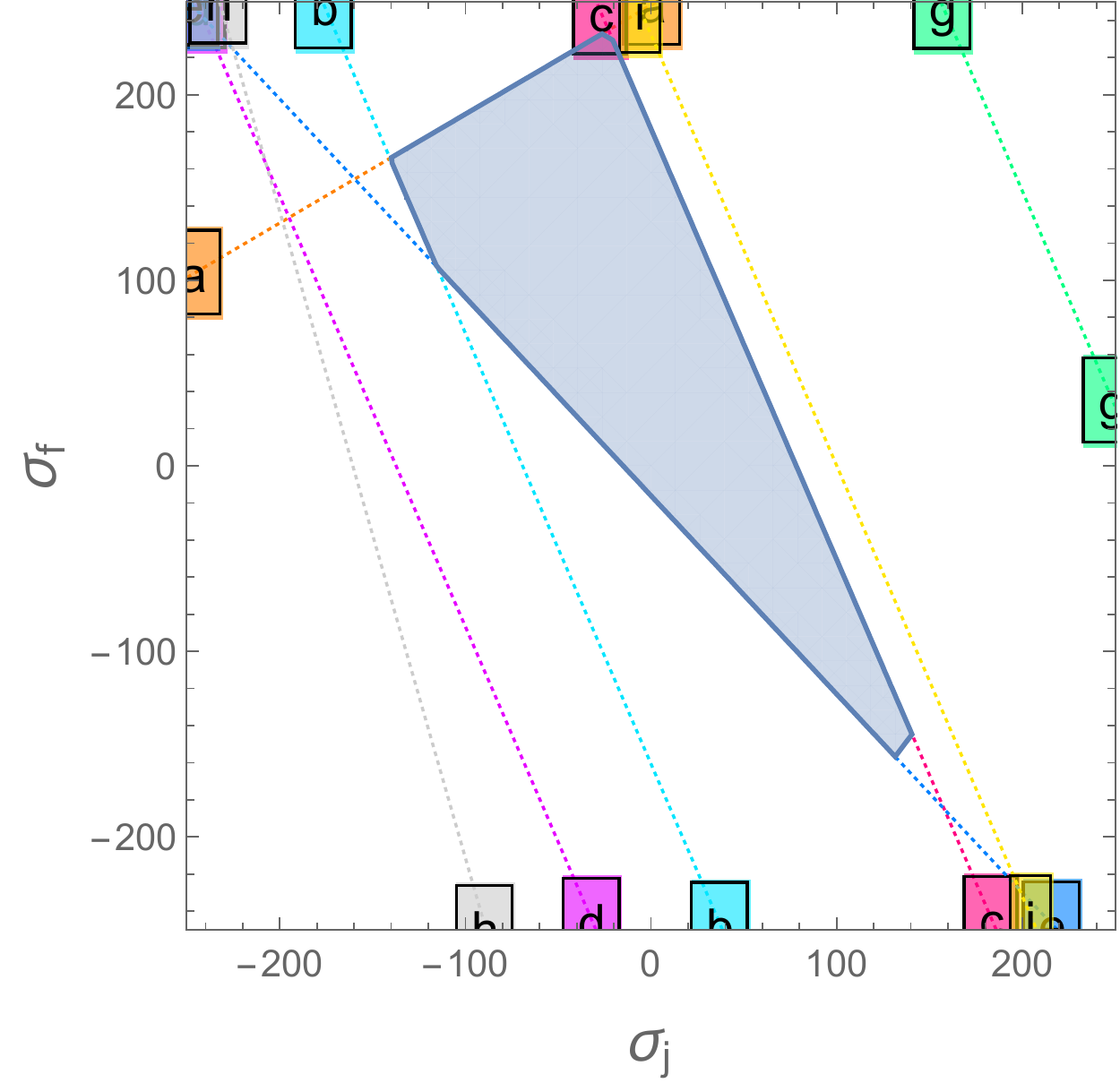}}
\caption{Various slices of the elastic domain of a magnesium single crystal, with stresses in MPa. Here, we use the notation
$\sigma_i = {\bf s}_i \cdot {\boldsymbol \sigma} {\bf n}_i$ for the resolved shear stress along the twin system
$i = \text{a}, \dots, \text{l}$ systems shown in Table \ref{tab:TwinInfo}.}
\label{fig:ys}
\end{figure}

Figure \ref{fig:ys} shows various two-dimensional slices through the yield surface.  Since the basal slip system is extremely easy, we see that it is dominant whenever there is any component of stress along it.  Figure \ref{fig:ys} (a) shows the section $\sigma_{13} - \sigma_a$ where we see the basal and the tension twin.   Notice that the yield surface is not symmetric about the origin reflecting the non-symmetry of twinning.  Also notice that scales on the two axes are different due to the ease of basal slip. Figure \ref{fig:ys}(b) shows that various modes can be closely competing in some loading directions. Figure \ref{fig:ys}(c-f) show that certain slices can be quite complex, with multiple twin modes participating to form the yield surface.

%%%%%%%%%%%%%%%%%%%%%%%%%%%%%%%%
%%%%%%%%%%%%%%%%%%%%%%%%%%%%%%%%
\section{Summary and conclusions}

\begin{algorithm}
\caption{Procedure for identifying possible and relevant twin modes in a given material of interest.}
\label{alg:TwinGenome}
\begin{algorithmic}[1]
	\Function{Calibrate Potential}{$\mathbf{e}_i$} \Comment{Identify stable parameters for interatomic potentials.}
		\For{Each interatomic potential}
			\State Identify a reasonable test range of $a$ and $c$.
			
			\For{Each test $a$ and test $c$}
				\State Compute energy of the crystal $E_\text{test}$.
			\EndFor
			
			\State $E_0^\text{pc}\gets \min_{a,c}E_\text{test}(a,c,\mathbf{e}_i)$.
			
			\State \textbf{Output:} $E_0^\text{pc}$ and $a,c$ which yield that value.
		\EndFor	
	\EndFunction
	\Function{Identify Possible Twins}{$a,c,\mathbf{e}_i$} \Comment{Identify all possible twin configurations.}
		\State Identify an admissible range of $\mu_i^{\ j}$.
		
		\State Compute the twinned lattice vectors $\mathbf{g}_i$.
		
		\State Compute deformation gradient $\mathbf{F}$ and right Cauchy-Green tensor $\mathbf{C}$.
		
		\ForAll{$\mu_i^{\ j}$ forming $\mathbf{C}$ satisfying Equation \eqref{eq:EigenvaluesC}}
			\State Compute the twinning shear $\mathbf{s}$ using Equation \eqref{eq:TwinningShear}.
			
			\State Compute the twinning normal $\mathbf{\hat{n}}$ using Equation \eqref{eq:TwinningNormal}.
		
			\State Compute the associated rotation matrix $\mathbf{Q}$ using Equation \eqref{eq:Q}.
		\EndFor
		
		\State \textbf{Store:} $\mathbf{s,\hat{n}},\mathbf{g}_i$.
	\EndFunction
	
	\Function{Compute Twin Energies}{$\mathbf{e}_i,\mathbf{g}_i$} \Comment{Use atomistic simulations to calculate the energies.}
		\State Construct a simulation cell consistent with Figure \ref{fig:LatticeWithCell}.
	
		\ForAll{Interatomic potentials}
			\State Compute the energy of the sample, $E^\text{tw}$.
			
			\State Compute the twin interface energy $\gamma^\text{tw}$ using Equation \eqref{eq:GammaTwin}.
		\EndFor
		
		\State \textbf{Store:} $\gamma^\text{tw},\Delta E^\text{max},\boldsymbol{\sigma}^\text{max}$
		
	\EndFunction
	
	\Function{Identify Relevant Twins}{$\mathbf{s,\hat{n}},\boldsymbol{\sigma}^\text{max}$} \Comment{Find twin modes which can be exploited.}
	
		\ForAll{Twin modes with $\gamma^\text{tw}$ and shear magnitude $s$ comparable to classical modes}
			\State Perform a nudged elastic band simulation calculate the energetic barrier to formation $\Delta E^\text{max}$.
					
			\State Calculate the attempt frequency $\nu_0$ using Equation \eqref{eq:AttemptFrequency}.
					
			\State Compute the Arrhenius barrier $K$ using Equation \eqref{eq:ArrheniusRelation}.
					
			\State Export the virial stresses at the maximum barrier $\boldsymbol{\sigma}^\text{max}$.
			
			\State Compute the resolved shear stress $\sigma_Y^i$ using Equation \eqref{eq:CRSS}.
						
			\State Form the elastic domain $\mathcal{Y}$ using Equation \eqref{eq:ElasticDomain}.
		\EndFor
		
		\State \textbf{Store:} All information about twin modes which lie on the yield surface.
	\EndFunction
\end{algorithmic}
\end{algorithm}

In this work, we have proposed a systematic approach to identifying the twin modes that are relevant to the deformation process of materials.  We combine the fundamental kinematic definition of a twin with atomistic methods of exploring the energy landscape to identify relevant twin modes with no empirical guidance.   Our approach is summarized in Algorithm \ref{alg:TwinGenome}.

We have applied this approach to magnesium.  Our framework shows that there are a very large number of twinning modes that are important in the deformation process of magnesium.  These results are consistent with the anomalous observations reported in the literature.  Our result argues that the physics of deformation in HCP materials is governed by an energetic and kinetic competition between numerous possibilities. Consequently, our findings suggest that the commonly used models of deformation physics need to be revisited in order to take into account a broader and richer variety of twin modes, and potentially points to new avenues of improving the mechanical properties.

In this work, we used the modified MEAM potential of Wu \emph{et al.} \cite{WuFrancisCurtin2015}.  Like all atomistic potentials, this one reproduces the energetics of configurations to which these were fitted but gives errors away from these configurations.     We have repeated some aspects of this work with potentials (the EAM potential of Sun \emph{et al.} \cite{SunEtAl2006},  MEAM potential of Kim \emph{et al.}\cite{KimEtAl2009} as well as an \emph{ab inito} electronic structure method (MacroDFT \cite{Phanish2012,PongaBhattacharyaOrtiz2015}).   We find differences.  However, the central observation -- that a large number of modes are important -- remains true independent of the potential use.  

Similarly, we have had to make choices in the various windows we have used for expediency.  These include the range of $\mu_i^{\ j}$, the number of super cells as well as the range of shear and twin boundary energy.  We have checked that our choices are relatively robust.  Still, it is possible that enlarging these choices significantly may lead to different modes.  The central observation -- that a large number of modes are important -- remains true independent of the choice.

{We conclude by drawing some comparisons between the results we have obtained as a part of this investigation, and some of the experimental observations of anomalous twin modes that were discussed in the introduction. Referencing Table \ref{tab:TwinInfo}, we see that some of the modes that we obtain are near values that have been experimentally observed. For instance, mode f is reasonably close to the $\{10\bar{1}3 \}\langle30\bar{3}2 \rangle$ twin observed in \cite{BrownEtAl2005,JiangEtAl2006,KitaharaEtAl2007,ReedHill1960,YoshinagaHoriuchi1963,YoshinagaObaraMorozumi1973,ReedHillRobertson1957b} and others. Likewise the discovery in this work of the relevance of a multitude of twin modes can potentially provide some insight on some of the experimentally-anomalous twin modes that have been observed, such as in \cite{MolodovEtAl2016}. Based on our findings, we conclude that these non-classical twin modes that have been widely observed in experimental literature are not anomalous readings, but, instead, are relevant twin modes that are being observed as a natural course of yielding in magnesium. }

{It should be noted that the framework we have proposed can be applied in a very broad fashion; extension to materials beyond magnesium or even hcp materials is only a trite calculation, and work is currently under way to show the application of this framework to 2D lattices. An additional area of ongoing research is the improvement of magnesium properties by alloying; in a forthcoming work, we shall use the framework developed here to study alloying and its effects on twins.}

\section{Acknowledgements}

This work draws from the doctoral thesis of DS at the California Institute of Technology, and was initiated when MP also held a position there.  Research was sponsored by the Army Research Laboratory and was accomplished under Cooperative Agreement Number W911NF-12-2-0022 and also through the National Defense Science and Engineering Graduate Fellowship. The views and conclusions contained in this document are those of the authors and should not be interpreted as representing the official policies, either expressed or implied, of the Army Research Laboratory or the U.S. Government. The U.S. Government is authorized to reproduce and distribute reprints for Government purposes notwith- standing any copyright notation herein. 
We gratefully acknowledge the support from the Natural Sciences and Engineering Research Council of Canada (NSERC) through the Discovery Grant under Award Application Number RGPIN-2016-06114 and the support of Compute Canada through the Westgrid consortium for giving access to the supercomputer grid. This research used resources of the Argonne Leadership Computing Facility, which is a DOE Office of Science User Facility supported under Contract DE-AC02-06CH11357.

%%%%%%%%%%%%%%%%%%%%%%%%%%%%%%
\bibliographystyle{plain}
\bibliography{Bibliography}

%Bibliography
%\printbibliography
%{\color{blue}  Need to format the citations uniformly.  Initials instead of first names, Journal names not capitals....}

\end{document}